\documentclass[]{spie}   
\usepackage{graphicx}
\usepackage{subfigure,amsmath}
\usepackage{booktabs}
\usepackage[suffix=]{epstopdf}

\newcommand{\eye}{\mathcal I}

\newcommand{\e}[1]{\times 10^{#1}}

\title{Kalman Filter Estimation for Focal Plane Wavefront Correction}
\author{Tyler D. Groff\supit{a}, N. Jeremy Kasdin\supit{a}
\skiplinehalf
\supit{a}Princeton University, Princeton, NJ USA
}

\authorinfo{Send correspondence to Tyler Groff
\texttt{tgroff@princeton.edu}}

\begin{document}
\maketitle

\begin{abstract} 
Space-based coronagraphs for future earth-like planet detection will require focal plane wavefront control techniques to achieve the necessary contrast levels. These correction algorithms are iterative and the control methods require an estimate of the electric field at the science camera, which requires nearly all of the images taken for the correction. We demonstrate a Kalman filter estimator that uses prior knowledge to create the estimate of the electric field, dramatically reducing the number of exposures required to estimate the image plane electric field. In addition to a significant reduction in exposures, we discuss the relative merit of this algorithm to other estimation schemes, particularly in regard to estimate error and covariance. As part of the reduction in exposures we also discuss a novel approach to generating the diversity required for estimating the field in the image plane. This uses the stroke minimization control algorithm to choose the probe shapes on the deformable mirrors, adding a degree of optimality to the problem and once again reducing the total number of exposures required for correction. Choosing probe shapes has been largely unexplored up to this point and is critical to producing a well posed set of measurements for the estimate. Ultimately the filter will lead to an adaptive algorithm which can estimate physical parameters in the laboratory and optimize estimation.
\end{abstract}


\section{Introduction}\label{sec:intro}
The desire to directly image extrasolar terrestrial planets has motivated much research into space-based missions. One approach proposed for direct imaging in visible to near-infrared light is a coronagraph, which use internal masks and stops to change the point spread function of the telescope, creating regions in the image of high contrast where a dim planet can be seen.  Coronagraphs possess an extreme sensitivity  to wavefront aberrations generated by the errors in the system optics (occulters are immune to this problem because the starlight never enters the telescope).  This necessitates wavefront control algorithms to correct for the aberrations and relax manufacturing tolerances and stability requirements within the observatory. In this paper we discuss the challenges associated with wavefront estimation and control in a coronagraphic imager. Advances in these correction algorithms have primarily been focused on development of the controller, by choosing some criterion that decides how best to suppress aberrations given an estimate of the electric field at that point in time. These estimators do not utilize prior knowledge of the electric field estimate, and as such require a large number of images to reconstruct the estimate. By utilizing prior estimates and the control history we develop a method that requires fewer images to update the estimate, thus improving the efficiency of the correction algorithm.

\section{Experimental Setup: Princeton High Contrast Imaging Laboratory}\label{HCIL}

\begin{figure}[ht!]
\centering
\includegraphics[width = .55\textwidth]{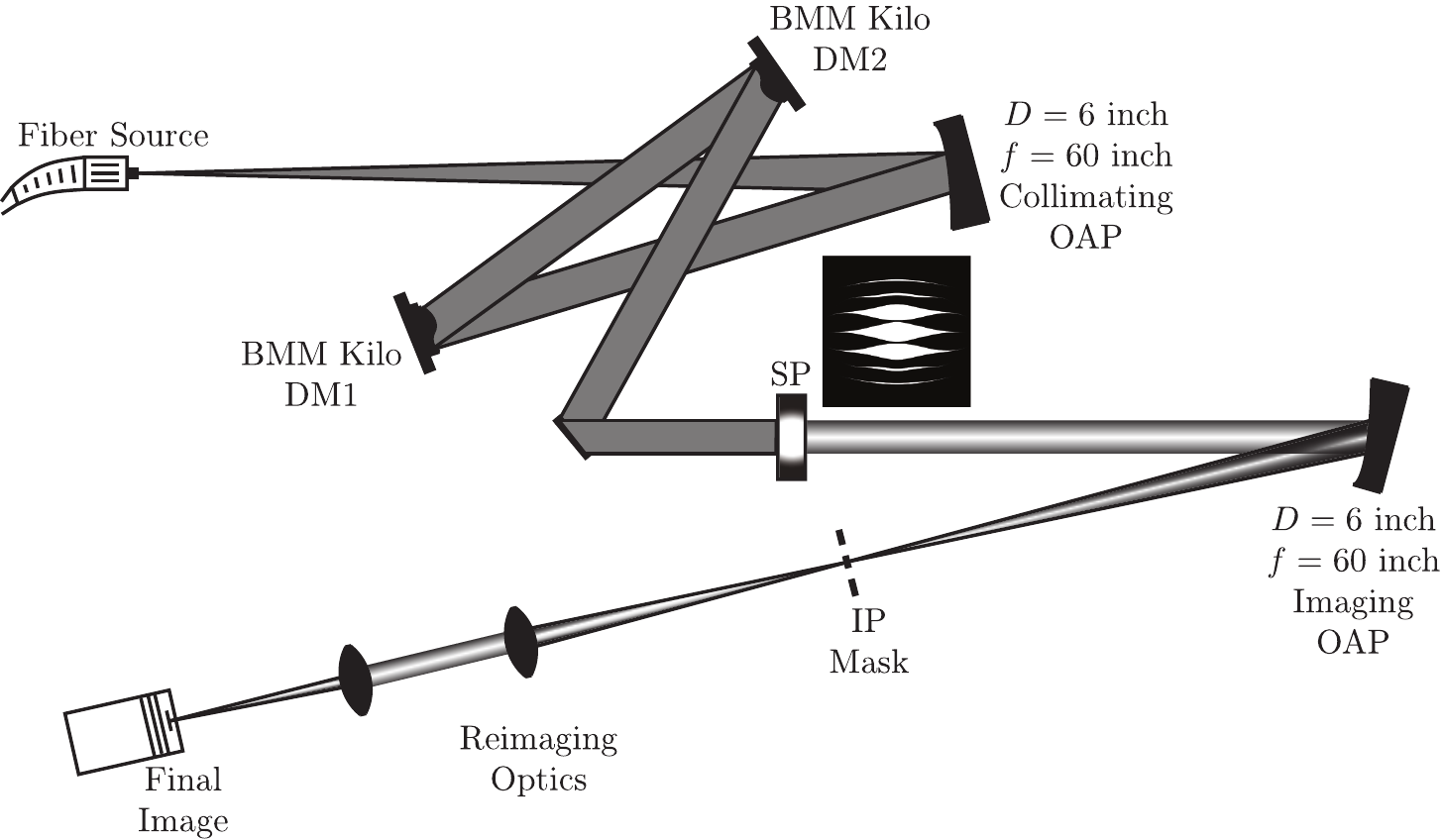}
\caption{Optical Layout of the Princeton HCIL.  Collimated light is incident on two DMs in series, which propagates through a Shaped Pupil, the core of the PSF is removed with an image plane mask, and the $90^\circ$ search areas are reimaged on the final camera.}
\label{optical}
\end{figure}

\begin{figure}[ht!]
\centering
\subfigure[] {
    \label{SP}
    \includegraphics[width = .205\columnwidth,clip=true,trim=.05in 0in .35in 0in]{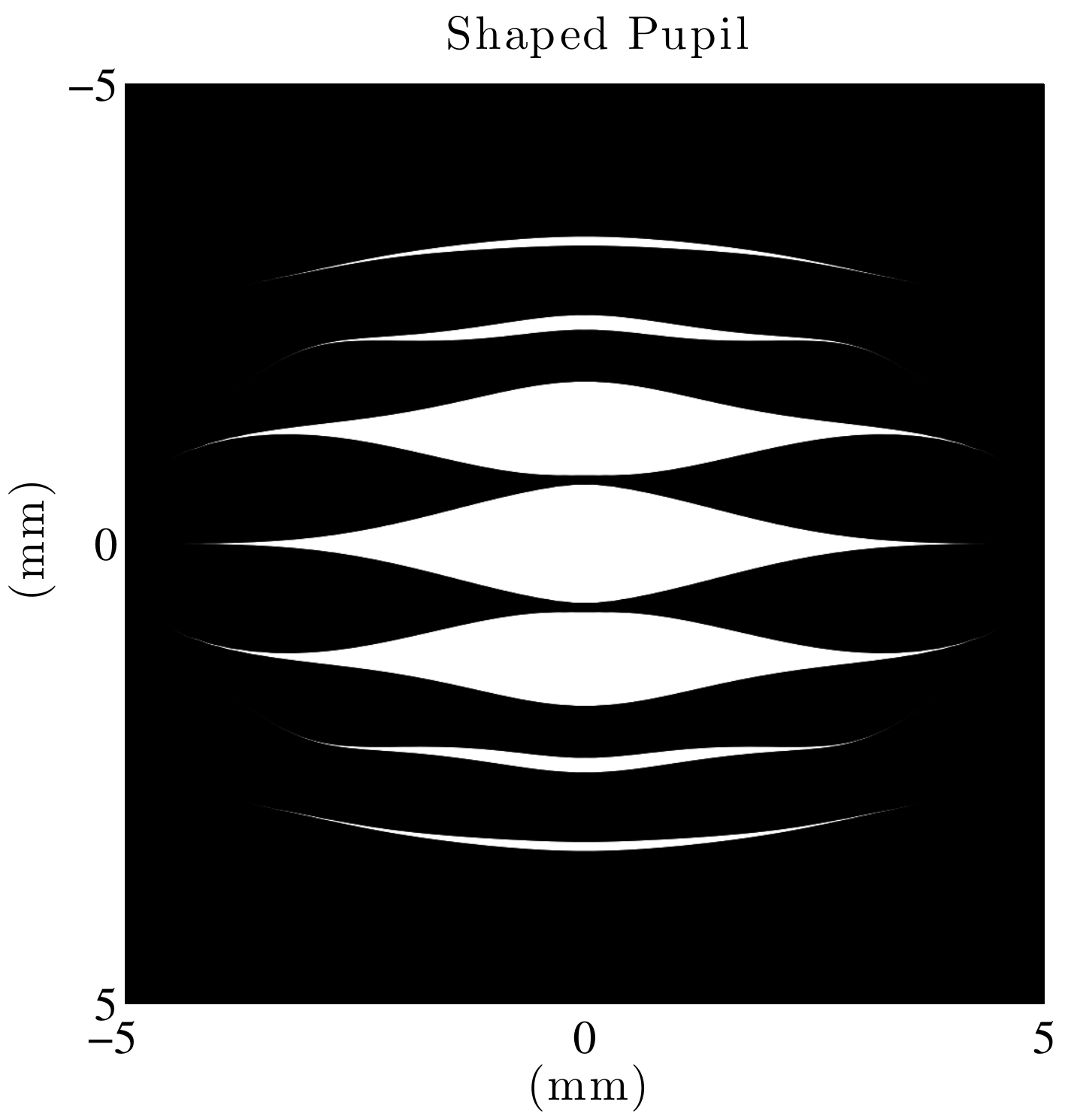}}
\subfigure[] {
    \label{Ideal_PSF}
    \includegraphics[width = .24\columnwidth,clip=true,trim=.05in 0in .35in 0in]{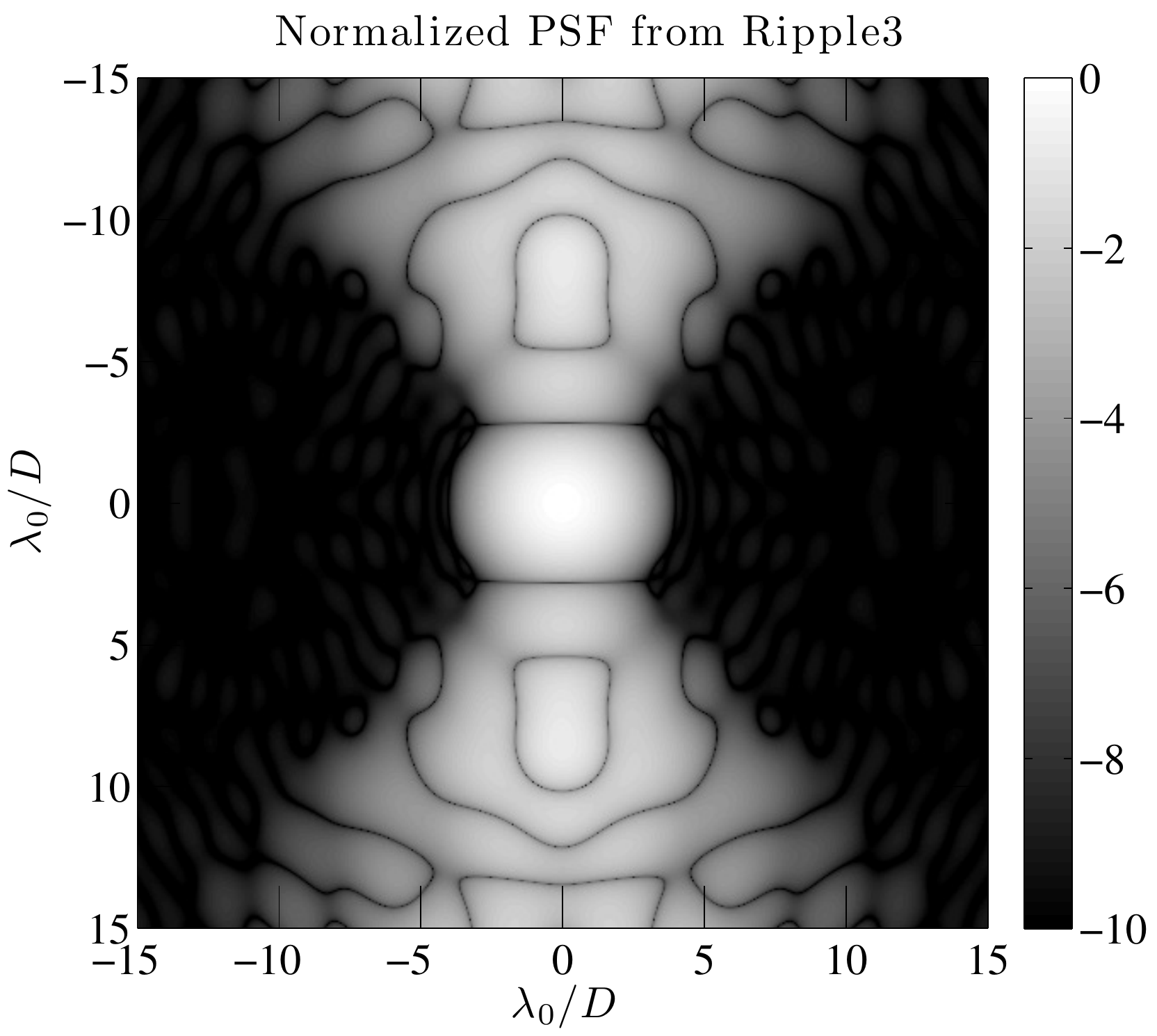}}
   \subfigure[]{
    \label{ab_PSF}
    \includegraphics[width = .205\columnwidth,clip=true,trim=.05in 0in .35in 0in]{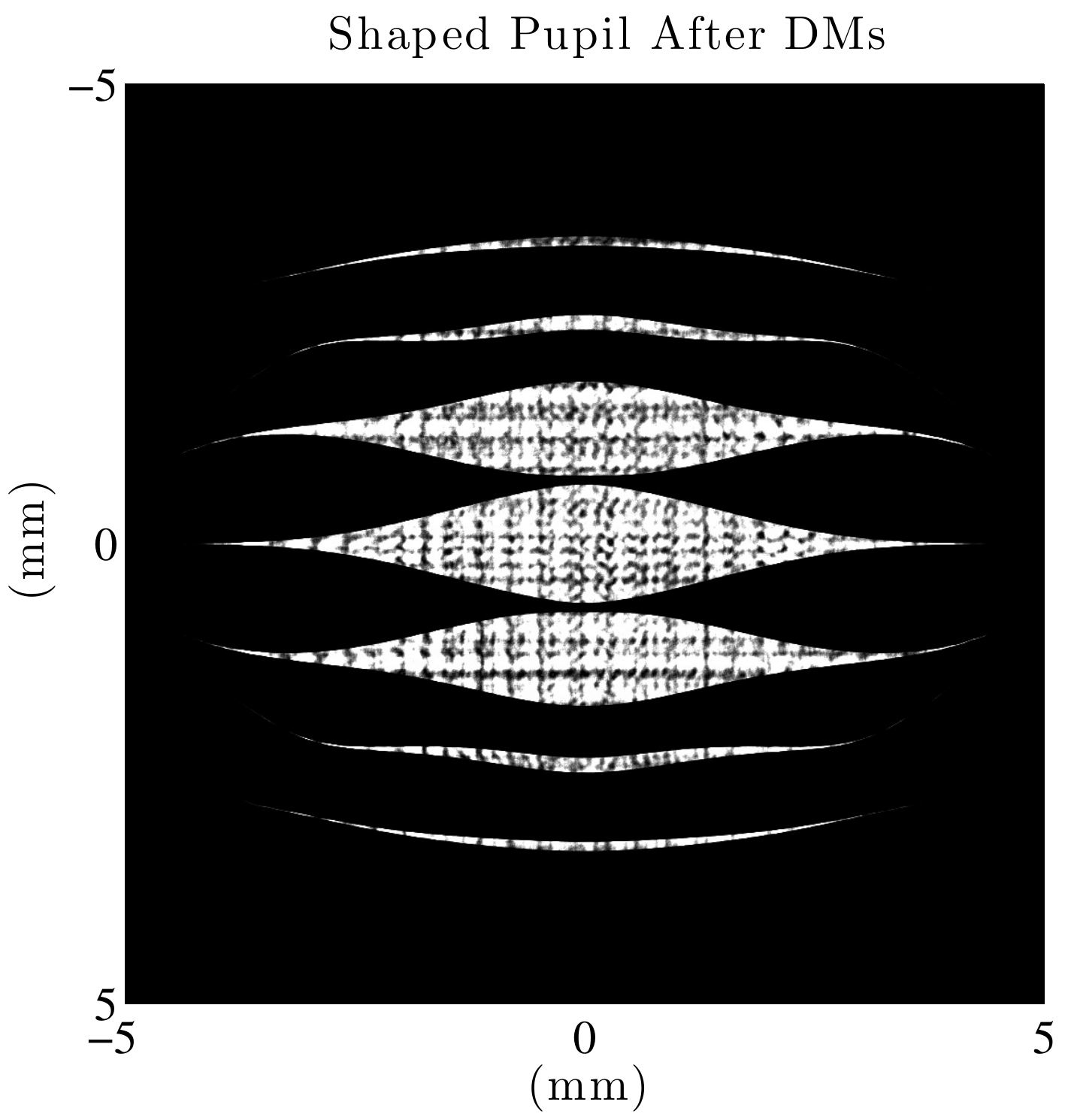}\label{SPab}}
    \subfigure[]{
    \label{ab_PSF}
    \includegraphics[width = .24\columnwidth,clip=true,trim=.05in 0in .35in 0in]{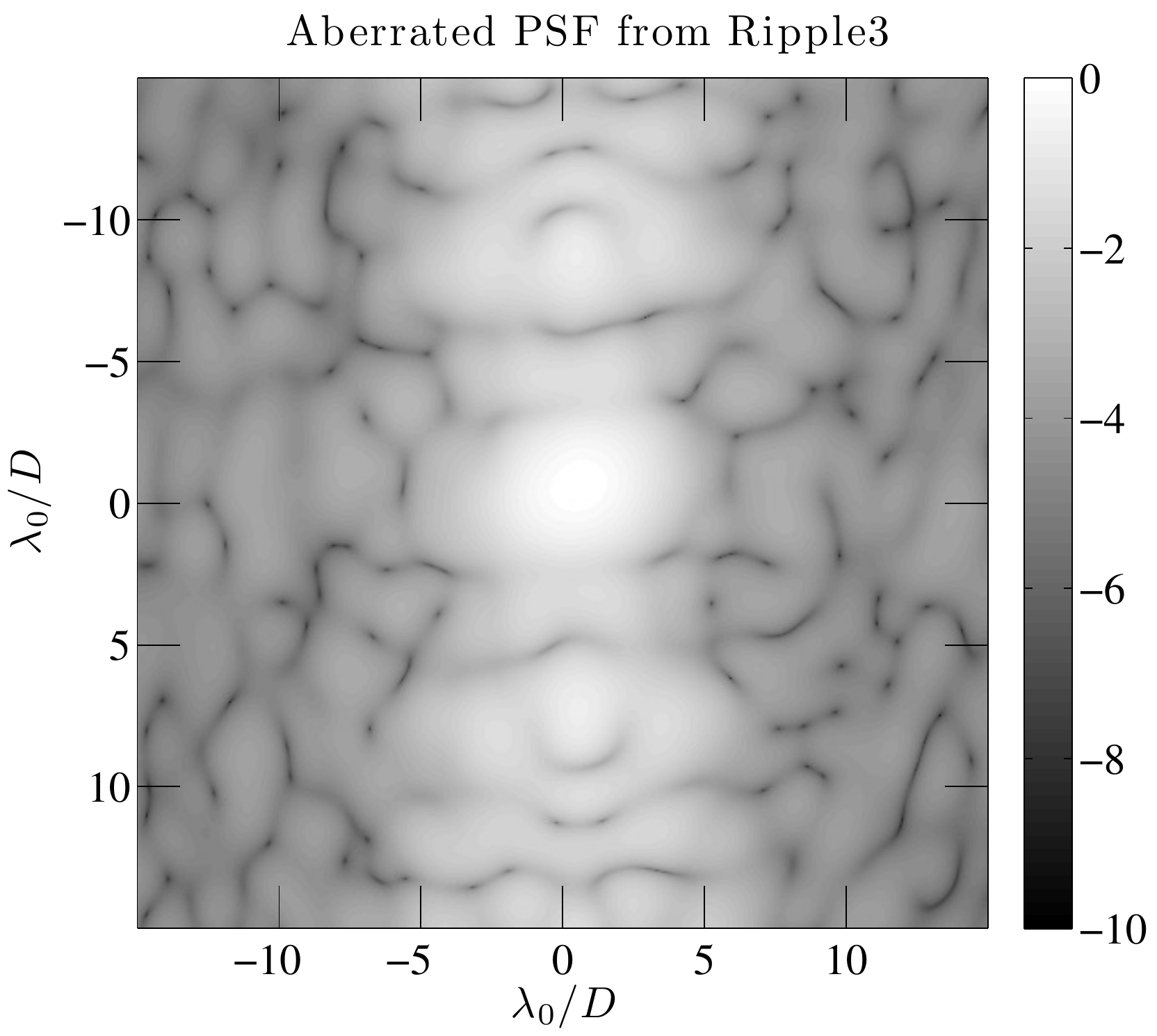}}
    \caption{(a) A shaped pupil.  (b)  The ideal PSF from a system using a shaped-pupil coronagraph.  (c)  Shaped Pupil with aberrations generated by Fresnel propagating the measured nominal shapes of the DMs to the pupil plane. Other sources of aberrations are not included because they have not been measured. (d) The PSF of the shaped pupil with the simulated aberrations.   The figures are in a log scale, and the log of contrast is shown in the colorbars.}
    \label{image}
\end{figure}

The High Contrast Imaging Laboratory (HCIL) at Princeton tests coronagraphs and wavefront control algorithms for quasi-static speckle suppression. The collimating optic is a six inch off-axis parabola (OAP) followed by two deformable mirrors (DMs) in series and a shaped pupil coronagraph, which is imaged with a second six inch OAP (Figure \ref{optical}).  We use a shaped pupil coronagraph, shown in Figure \ref{SP}, and described in detail in Belikov et al.\cite{belikov2007broadband}.  This coronagraph produces a discovery space with a theoretical contrast of $3.3\times10^{-10}$ in two $90^\circ$ regions as shown in Figure \ref{Ideal_PSF}. At the Princeton HCIL, the  aberrations in the system result in an uncorrected average contrast of just under $1\times10^{-4}$ in the area immediately surrounding the core of the point spread function (PSF), which agrees with the simulations shown in Figure \ref{ab_PSF}. Since the coronagraph is a binary mask, its contrast performance is fundamentally achromatic, subject only to the physical scaling of the PSF with wavelength.

\section{Stroke Minimization in Monochromatic Light}\label{stroke}
Using a Taylor series expansion of the aberrated electric field it was shown in Pueyo et al. \cite{pueyo2009optimal} that, to first order, two DMs in series are capable of correcting both amplitude and phase aberrations, resulting in symmetric dark holes in the image plane. Physically, such a controller relies on the amplitude-to-phase mixing resulting from propagation of the field between  non-conjugate planes (the first DM to the second).  If the magnitude of the first DM's actuation is chosen correctly, then it can exactly conjugate the amplitude variations at the pupil plane at which point the second DM, assumed to be at a conjugate pupil, can correct the residual phase aberrations. Algorithmically this is achieved through a number of methods, each with its set of advantages and disadvantages. In general, focal plane wavefront correction methods are broken down into a wavefront estimation step followed by a control step where the correction is decided. In this paper we use the stroke minimization algorithm as the controller to test the estimator. It corrects the wavefront by minimizing the actuator stroke on the DMs subject to a target contrast value\cite{pueyo2009optimal}.  Expressing the DM actuator amplitudes as a vector, $u$, the optimization problem can be written as
\begin{equation}
\begin{array}{ccc}
\mbox{minimize} & & \sum_{k=1}^N a_k^2 = u^Tu \medskip\\
\mbox{subject to} & & I_{DZ} \le 10^{-C},
\end{array}
\end{equation}
where $a_k$ is the commanded height of actuator $k$, $I_{DZ}$ is the residual intensity in the dark hole after correction, and $C$ is the target contrast.  We solve the optimization by approximating $I_{DZ}$ as a quadratic form,

\begin{equation}
I_{DZ} \cong \left(\frac{2\pi}{\lambda}\right)^2u^TMu + \frac{4\pi}{\lambda} \Im \{b^T\} u+d
\end{equation}
where $b$ is a vector describing the interaction between the DM shape and the aberrated field, $d$ is a vector that expresses contrast in the dark hole, and $M$ is the matrix which describes the linearized mapping of DM actuation to intensity in the dark hole.  The resulting quadratic subprogram is easily solved by augmenting the cost function via Lagrange multiplier, $\mu$, and solving for the commanded actuator heights:
\begin{align}
J &= u^T\left( \eye + \mu_0\frac{4 \pi^2}{\lambda_0^2} M_0\right)u + \mu_0 \frac{4 \pi}{\lambda_0} u^T \Im\{b_0\} + \mu_0 \left(d_0 - 10^{-C}\right). \label{eq:monocost}\\
u_{opt} &= - \mu_0 \left(\frac{\lambda_0}{2\pi} \eye + \mu_0 \frac{2\pi}{\lambda_0} M_0 \right)^{-1} \Im\{b_0\}.\label{eq:monocontrol}
\end{align}

We find the optimal actuator commands via a line search on $\mu$ to minimize the augmented cost function (Eq.~(\ref{eq:monocost})). Since this is a quadratic subprogram of the full nonlinear problem we can iterate to reach any target contrast \cite{pueyo2009optimal}. In addition to regularizing the problem of minimizing the contrast in the search area, minimizing the stroke has the added advantage of keeping the actuation small and thus within the linear approximation. If the DM model and its transformation to the electric field (embedded in the $M$ matrix) were perfectly known, the achievable monochromatic contrast using stroke minimization would be limited only by estimation error as long as the DM actuation remains within the bounds of the linearization.  Our ability to estimate the field is driven  largely by the residual model error associated with the DM. 

\section{Least-Squares Estimation Using Pairwise Measurements}\label{batch}
To date, almost all high-contrast wavefront control approaches use DM-Diversity\cite{giveon2007broadband} to estimate the wavefront. We thus use it as a baseline to compare with the Kalman filter estimation scheme. It has given the best results to date using the stroke minimization control algorithm at the Princeton HCIL in both monochromatic and broadband suppression \cite{pueyo2009optimal,groff2010progress}. The linearized interaction of the DM actuation and the aberrated electric field can be written in matrix form by taking difference images using $j$ pre-determined shapes with amplitudes prescribed by the normalized intensity of the aberrated field \cite{borde2006speckle,giveon2007broadband}. The image $I_j^+$ is taken with one deformable mirror shape, $\phi_j$, while $I_j^-$ is the image taken with the negative of that shape, $-\phi_j$, applied to the deformable mirror. The difference of each conjugate pair is then used to construct a vector of noisy measurements, 
\begin{equation}
z = \left[\begin{array}{c}I_1^+ - I_1^-\\ \dots \\ I_j^+ - I_j^-\end{array}\right],
\end{equation}
for each pixel. By making the measurement the difference of two images we can remove incoherent light, such as detector noise and planet light, leaving only the coherent component of the field. Additionally, since the images were taken with conjugate deformable mirror shapes, only the contribution of the product of the DM field and the aberrated field to the intensity measurement exists in $z$. Defining $x$ as the image plane electric field state, we write $z$ as a linear equation in $x$ and include additive noise, $n$,
\begin{equation}
z = H x + n
\end{equation}
where $H$ is the observation matrix that relates the observed quantity to the state we seek to estimate. By writing $x$ as the real and imaginary parts of the electric field at a specific pixel,
\begin{equation}
x = \left[\begin{array}{c} \Re \{C\{A g\}\} \\ \Im\{C\{Ag\}\} \end{array}\right],
\end{equation}
we can construct the observation matrix, $H$, so that it contains the real and imaginary parts of the $j^{th}$ DM perturbation, $C\{A\phi_j\}$, in each row. Taking multiple measurments, the observation matrix is given by
\begin{equation}
H = 4 \left[\begin{array}{cc}\Re\{C\{A \phi_1 \}\} & \Im\{C\{A \phi_1 \}\}\\ \vdots&\vdots \\ \Re\{C\{A \phi_j\} \} & \Im\{C\{A \phi_j\} \} \end{array}\right].
\end{equation}
The product $H x$ will then match the intensity distribution in the measurement $z$. With at least three measurements, $j\geq3$, we can take a left pseudo-inverse to solve for the estimate of the real and imaginary parts of the aberrated field at each pixel in the image plane with least-squares minimal error:
\begin{equation}
\hat x = (H^TH)^{-1}H^T z.
\end{equation}

While the problem is still invertible using two image pairs to construct $z$ and $H$, a minimum of 3 image pairs must be used to create an overdetermined system that will produce a unique estimate with least-squares minimal error from the available data. Practically, we find that 4 image pairs must be used to get a good enough estimate at the Princeton HCIL. Consequently, 8 images are taken per iteration to estimate the electric field when using the DM diversity algorithm.

\begin{figure}[h]
\centering
\subfigure[]{\includegraphics[width = 0.26\paperwidth,clip=true,trim=.05in 0in .35in 0in]{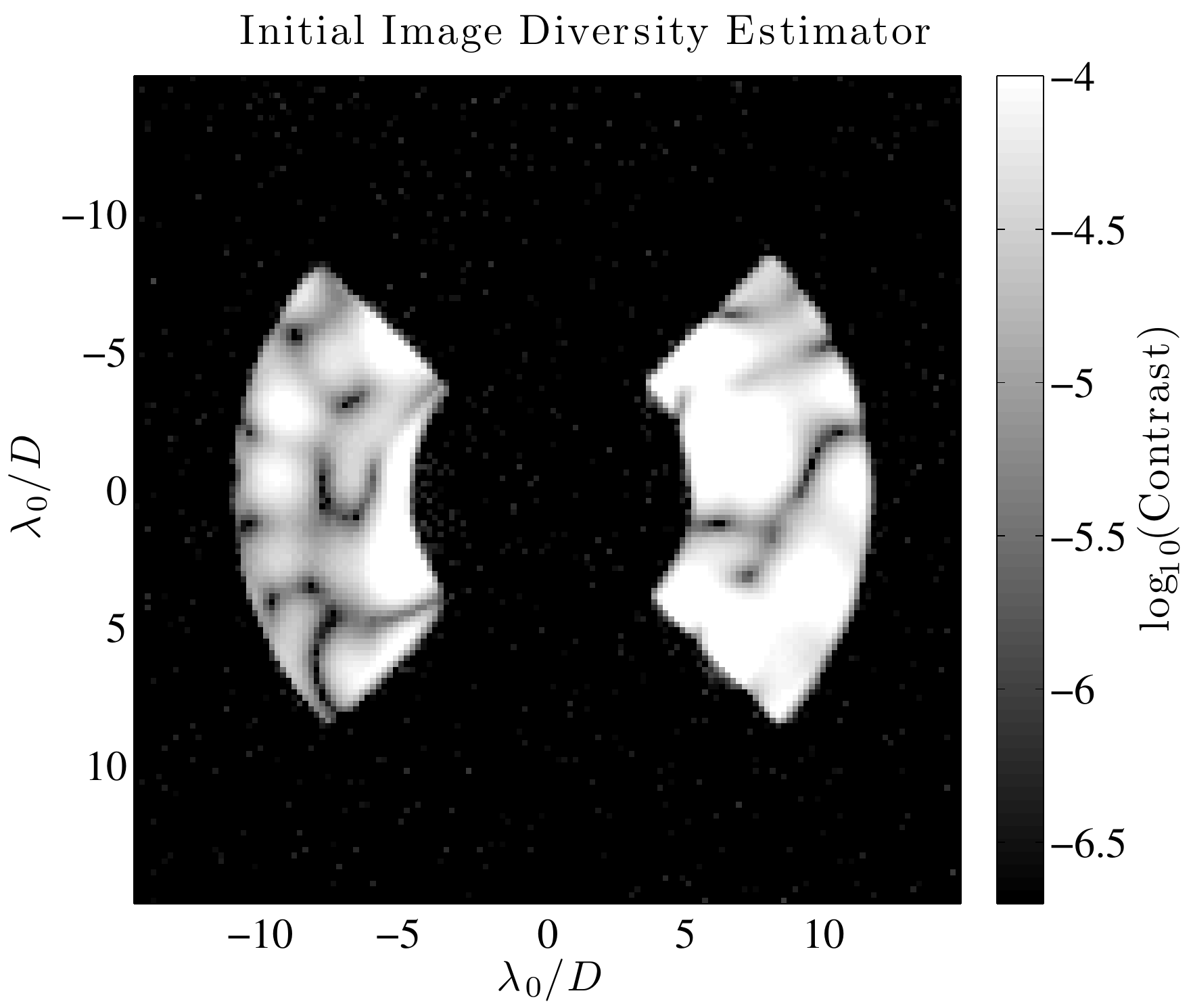}\label{mono_initial}}
\subfigure[]{\includegraphics[width = 0.25\paperwidth,clip=true,trim=.05in 0in .5in 0in]{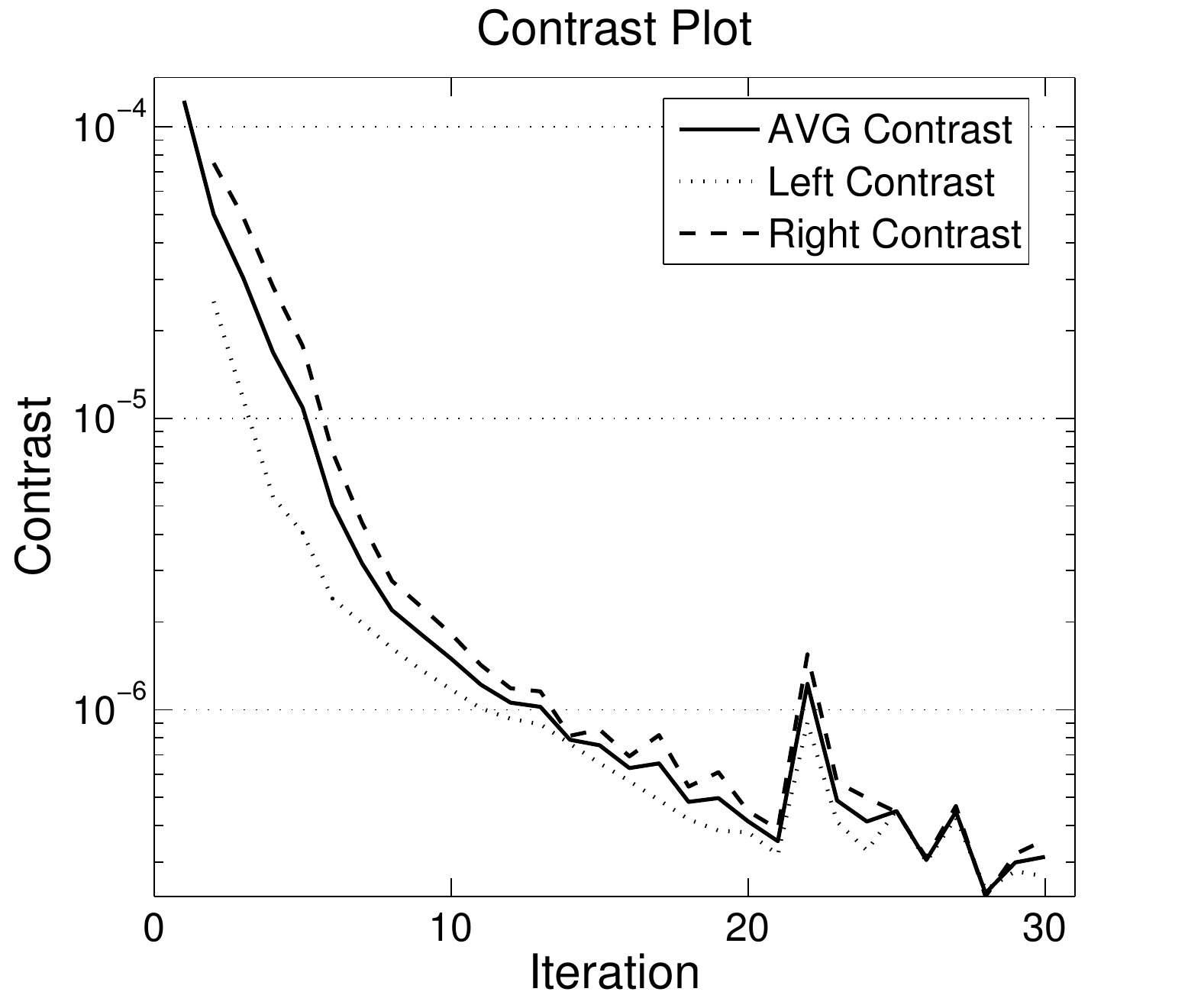}\label{mono_contrast}}
\subfigure[]{\includegraphics[width = 0.26\paperwidth,clip=true,trim=.05in 0in .35in 0in]{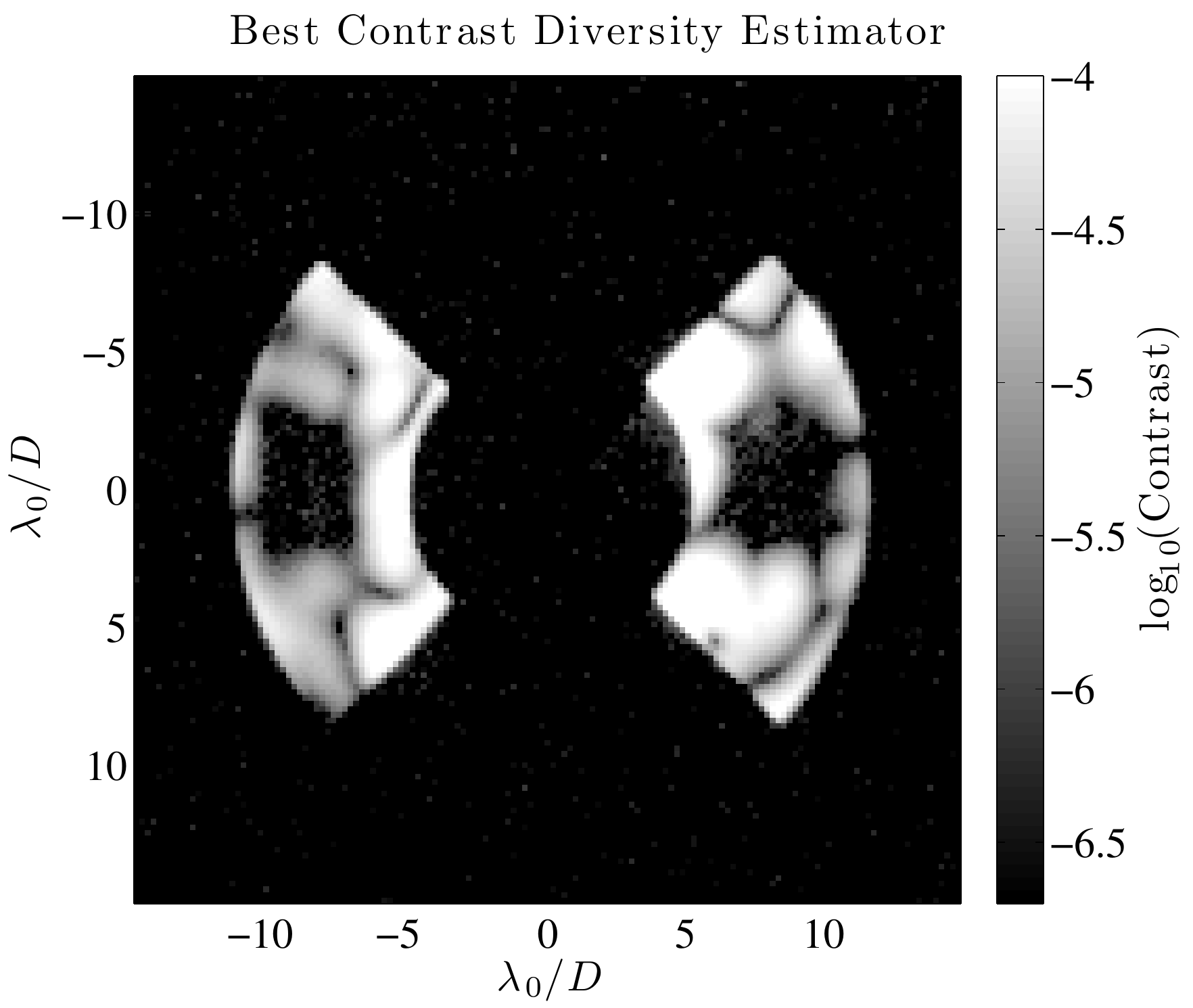}\label{mono_corrected}}
\caption{Experimental results of sequential DM correction using the DM-Diversity estimation algorithm.  The dark hole is a square opening from 7--10 x -2--2 $\lambda/D$ on both sides of the image plane.  (a) The aberrated image.  (b)  Contrast plot. (c) The corrected image. Image units are log(contrast). }\label{diversity}
\end{figure}

The laboratory starts at an initial contrast of $1.23\times10^{-4}$ (Fig.~\ref{mono_initial}). Using the least-squares estimation technique it is capable of reaching an average contrast of $2.3\times10^{-7}$ in a (7-10)x(-2-2) $\lambda/D$ region within 30 iterations (Fig.~\ref{mono_corrected}) on both sides of the image plane, a unique capability that is a result of the two deformable mirrors in the system. In 20 iterations of the algorithm, requiring a total of 160 estimation exposures, the system reached a contrast level of $3.5 \times 10^{-7}$.

\section{Constructing the Kalman Filter Estimator}\label{sec:filter}
\begin{figure}[ht]
\centering
\includegraphics[width = 0.6\textwidth]{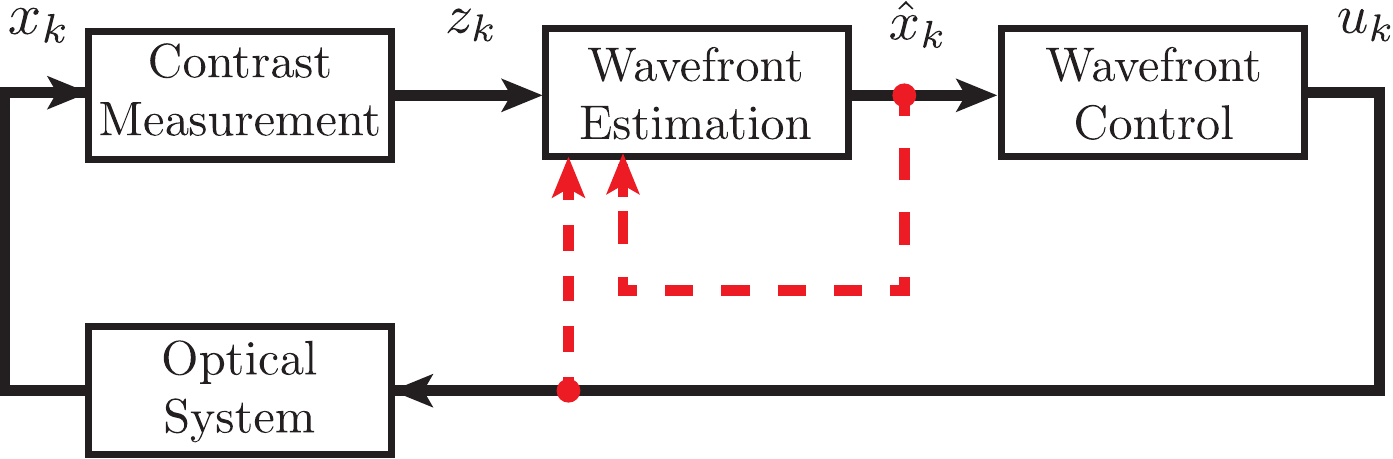}
\caption[Feedback Block Diagram]{Block diagram of a standard FPWC control loop. At any time step, $k$, only the intensity measurements, $z_k$, provide any feedback to estimate the current state, $x_k$, for control. The red dashed lines show additional feedback from the prior electric field (or state) estimate, $\hat x_k$, and the control signal, $u_k$, used to suppress it. }\label{fig:feedback}
\end{figure}

The DM-Diversity algorithm \cite{giveon2011pair} is quite effective, but it is limited by the fact that it is only a batch process method. As shown in Fig.~\ref{fig:feedback}, it does not close the loop on the state estimate. Therefore all state estimate information, $\hat x$, acquired about the electric field in the prior control step is lost. Thus each iteration operates as if it was the first, requiring that we take a full set of estimation images to estimate the field again. 
In addition to being very costly with regard to exposures, the measurements will become progressively noisier as higher contrast levels are reached. If we include feedback of the state estimate we will have a certain degree of robustness to new, noisy measurements by including information from prior measurements with better signal-to-noise. Since we already have demonstrated a model based controller, we should be able to use this model to predict the change in the electric field after the controller has applied a DM command. In doing so we do want to consider the relative effect of process and detector noise to optimally combine an extrapolation of the state estimate with new measurement updates. This is exactly the problem a discrete time Kalman filter solves.

With the Kalman filter we seek to optimally include new measurements to improve the state and covariance estimates. These noisy measurements, $z_k = y_k + n_k$, will still be difference images of probe pairs. The conjugate pairs allow us to construct a linear observation matrix, $H_k$. If we were not in a low aberration regime our observer would have to be nonlinear. This is not impossible for a Kalman filter, but can make it highly biased \cite{stengel1994optimal} and computationally expensive. As we decide how to optimally update the field, we must also have an estimate of our measurement noise covariance, which we define as
\begin{equation}
R_k = E[n_kn_k^T].
\end{equation}

Like the DM Diversity estimator \cite{giveon2007broadband,giveon2011pair}, the Kalman filter produces an estimate with least-squares minimal error. Since the Kalman filter operates on the estimate in closed loop, the weighted cost function used to derive the solution for the DM Diversity estimator,
\begin{equation}
J = \frac{1}{2} [H_k\hat x_k - z_k]^T R_k^{-1}  [H_k\hat x_k - z_k],\label{eq:batchcost}
\end{equation}
will not adequately represent the error contributions in the system. We must also include an estimate of the state covariance, $P_k(-)$, since this will also propagate error in the estimate update. Defining the error as both the difference between the noisy observation and the estimated observation, $H_k\hat x_k (+) -z $, and the difference between our current estimate and the estimate extrapolation, $(\hat x_k -  \hat x_k(-))$, we write the quadratic cost function as
\begin{equation}
J = \frac{1}{2} \left[\hat x_k - \hat x_k(-)]^T P_k(-)^{-1} [\hat x_k - \hat x_k(-)\right] + \frac{1}{2} \left[H_k\hat x_k - z_k\right]^T R_k^{-1}  \left[H_k\hat x_k - z_k\right].
\end{equation}
We can formulate the cost in matrix form as
\begin{align}
J &= \frac{1}{2}\begin{bmatrix} \hat x_k - \hat x_k(-) \\ H_k \hat x_k - z_k\end{bmatrix}^T \begin{bmatrix} P_k(-) & 0 \\ 0 & R_k \end{bmatrix}^{-1} \begin{bmatrix} \hat x_k - \hat x_k(-) \\ H_k \hat x_k - z_k\end{bmatrix} \\
&= \frac{1}{2} (\tilde H_k \hat x_k - \tilde z_k)^T \tilde R_k^{-1}  (\tilde H_k\hat x_k - \tilde z_k), \label{eq:kalmanquad}
\end{align}
where we have now defined a new set of augmented matrices as
\begin{align}
\tilde H_k &= \begin{bmatrix}\eye\\H_k\end{bmatrix}, \label{eq:augH} \\
\tilde z_k &= \begin{bmatrix}\hat x_k(-) \\ z_k \end{bmatrix},\\
\tilde R_k &= \begin{bmatrix} P_k(-) & 0 \\ 0 & R_k \end{bmatrix}. \label{eq:augR}
\end{align}
Evaluating the partial derivative of Eq.~\ref{eq:kalmanquad} the state estimate update is given by
\begin{equation}
\hat x_k(+)=  \hat x_k(-) + P_k(-) H_k^T\left[H_kP_k(-)H_k^T + R_k\right]^{-1} \left[z_k - H_k \hat x_k(-) \right]. \label{eq:deriveupdate}
\end{equation}
From Eq.~\ref{eq:deriveupdate}, we define the optimal gain to be
\begin{equation}
K_k = P_k(-) H_k^T[H_kP_k(-)H_k^T + R_k]^{-1}.\label{eq:gainderive}
\end{equation}

Eq.~\ref{eq:gainderive} optimally combines the prior estimate history with measurement updates to minimize the total error contributions based on the expected state and measurement covariance. Much like the batch process method the Kalman filter produces a solution that minimizes a quadratic cost function, Eq.~\ref{eq:batchcost}, but it is also subject to the constraining dynamic equations given by $\hat x_k(-)$ and $P_k(-)$. However, looking at Eq.~\ref{eq:kalmanquad} there is a major advantage of the Kalman filter in its minimization of the cost function. For $\tilde H_k$ to be overdetermined, we only require a single measurement. Thus, at a fundamental level the Kalman filter is formulated in such a way that it solves a least squares, left pseudo-inverse problem, regardless of the number of measurements taken. This gives us the freedom to minimize the number of exposures required to estimate the field to a precision adequate for suppressing the field to the target contrast level.

Finally, we update the state covariance estimate, $P_k(+)$, by applying Eq.~\ref{eq:deriveupdate} to the expected value of the covariance,
\begin{equation}
P_k(+) = E[(\hat x_k(+) - x_k)(\hat x_k(+) - x_k)^T],
\end{equation}
which gives
\begin{equation}
P_k(+) = [P_k(-)^{-1} + H_k^T R_k^{-1} H_k ]^{-1}.\label{eq:covderive}
\end{equation}
For the weighted form of the least-squares method, the covariance of the batch process method described in \S\ref{batch} is given by
\begin{align}
P &= E\left[(x-\hat x)(x-\hat x)^T\right]\\
&= \left(H^T R^{-1} H\right)^{-1}. \label{eq:batchcov}
\end{align}
As shown in Eq.~\ref{eq:batchcov}, the state covariance of the batch process method resets after every control step, and is tied to the noise in  that particular set of measurements. However, the covariance of the Kalman filter is also a function of the prior state covariance. Looking at Eq.~\ref{eq:covderive},  $H_k^T R_k^{-1} H_k$, is guaranteed to be positive definite. Thus additional measurements taken at each iteration will act to reduce the magnitude of the covariance since additional measurements can do nothing but make the inversion smaller. 

We can use the contrast normalization for the measurements to get an idea of the estimator's robustness. If we do not take a long enough exposure in the probe images $R_k$ will become large, indicating a poor signal to noise ratio. In this case the covariance may not get better, but it is guaranteed not to get worse. In the batch-process estimator, we are stuck with these measurements and will receive an estimate with large covariance. In this case the control will not be effective, which is why we often see jumps in contrast when using this estimator once we reach low contrast levels, as seen in Fig.~\ref{diversity}. In the case of the Kalman filter, this high covariance is dampened by the contribution of prior covariance estimates via $P_k(-)$, stabilizing the state estimate and its covariance in the event of a bad measurement. Since we cannot guarantee that a probe will provide good signal, particularly at low contrast levels, this is an extremely attractive component of the Kalman filter estimator.

To complete the filter we need to propagate the prior estimate, $x_{k-1}(+)$, to the current time step. The filter extrapolates to the current state estimate, $\hat x_k(-)$, by applying a time update to the prior state estimate via the state transition matrix, $\Phi_{k-1}$, and numerically propagating the control output from stroke minimization at the prior iteration, $u_{k-1}$,  via a linear transformation described by $\Gamma_{k-1}$. We also have a disturbance from the process noise, $w_{k-1}$, which is propagated to the current state of the electric field via the linear transformation, $\Lambda_{k-1}$. Assuming these components are additive, the state estimate extrapolation is
\begin{equation}
\hat x_k (-) = \Phi_{k-1} \hat x_{k-1}(+) + \Gamma_{k-1} u_{k-1} + \Lambda_{k-1} w_{k-1}. \label{eq:estpropproc}
\end{equation}
 We will apply the linearized optical model used to develop the batch process estimation method and stroke minimization control algorithm. Using a linearized model avoids generating arbitrary bias in the estimate at each pixel, a common problem with a nonlinear filter \cite{gelb1974optimal}. The first step in propagating the state forward in time is to update any dynamic variation between the discrete time steps with the state transition matrix, $\Phi_{k-1}$. In this system, $\Phi_{k-1}$ captures any variation of the field due to temperature fluctuations, vibration, or air turbulence that perturb the optical system. To simplify the model, we recognize that there is no reliable way to measure or approximate small changes in the optical system over time with alternate sensors; we assume that the state remains constant between control steps, making the state transition matrix, $\Phi_{k-1}$,  $\Phi_{k-1} = \Phi = \mathcal{I}$. Each submatrix for $\Gamma$, shown in Table~\ref{filtertable}, is of dimension $2 \times 2N_{DM}$ and represents the control effect on a single pixel of the matrix. Making the standard assumption that the process noise is gaussian white noise, the expected value of the state when we extrapolate is
 \begin{equation}
 \hat x_k (-) = \Phi_{k-1} \hat x_{k-1}(+) + \Gamma_{k-1} u_{k-1}. \label{eq:estprop}
\end{equation}
It's associated covariance extrapolation is then given by
\begin{equation}
P_k(-) = \Phi_{k-1} P_{k-1}(+) \Phi_{k-1}^T + Q_{k-1}. \label{eq:covprop}
\end{equation}

Combining Eq.~\ref{eq:deriveupdate}, Eq.~\ref{eq:gainderive}, and, Eq.~\ref{eq:covderive} with the extrapolation equations, this form of the filter consists of five equations that describe the state estimate extrapolation, covariance estimate extrapolation, filter gain computation, state estimate update, and covariance estimate update at the $k^{th}$ iteration \cite{stengel1994optimal}:
\begin{align}
\hat x_k (-) &= \Phi_{k-1} \hat x_{k-1}(+) + \Gamma_{k-1} u_{k-1}. \label{eq:estextrap}\\
P_k(-) &= \Phi_{k-1} P_{k-1}(+) \Phi_{k-1}^T + Q_{k-1} \label{eq:covextrap}\\
K_k &= P_k(-)H_k^T \left[H_kP_k(-)H_k^T + R_k \right]^{-1}\label{eq:gain}\\
\hat x_k(+) &= \hat x_k(-) + K_k \left[z_k - H_k \hat x_k(-) \right] \label{eq:estupdate}\\
P_k(+) &= \left[P_k(-)^{-1} + H_k^T R_k^{-1}H_k \right]^{-1}\label{eq:covupdate}
\end{align}

With $H_k$, $z_k$, $\Gamma$, $\hat x$, and $u_k$ constructed, the dimension and form of the rest of the filter follows. Table~\ref{filtertable} and Table~\ref{filtervectors} define all the matrices and vectors in the filter equations for this problem and provides their dimensionality for clarity. The initialization of the covariance, $P_0$, is critical for the performance of the filter. In our system this cannot be measured, so we must initialize with a reasonable guess. 
The focal plane measurements $z_k$ are identical to that of \S\ref{batch}, and are constructed into a vertical stack of difference images taken in a ``pair-wise" fashion to produce $j$ measurements for $n$ pixels. Likewise $H_k$ takes on a similar form, and is a matrix constructed from the effect of a specific deformable mirror shape $\phi_j$ on the real and imaginary parts of the electric field in the image plane. Finally, we compute the covariance update, $P_k(+)$, based on the added noise from the new measurements. The estimated state is a vertical stack of the real and imaginary parts of the electric field at each pixel of the dark hole in the image plane. The control signal $u$ is a vertical stack of the actuators of each DM, with $DM1$ being stacked on top of $DM2$. Since we are only considering process noise at the DMs, the process disturbance $w$ is a vertical stack of the variance expected from each actuator. 
 \begin{table}[ht!]
\centering
\vline
   \begin{tabular}{ cc } 
         \toprule
      Matrix & Dimension \\
      \midrule
      	$\Phi = \mathcal I $& $(2 \cdot N_{pixels}) \times (2 \cdot N_{pixels})$\bigskip\\
	$\Gamma  = \begin{bmatrix} 
	\begin{bmatrix} \Re \{G_{DM1}\} & \Re \{G_{DM2}\} \\  \Im \{G_{DM1}\} & \Im \{G_{DM2}\} \end{bmatrix}_1 \\
	\vdots \\
	\begin{bmatrix} \Re \{G_{DM1}\} & \Re \{G_{DM2}\} \\  \Im \{G_{DM1}\} & \Im \{G_{DM2}\} \end{bmatrix}_n
	\end{bmatrix}$ &  $(2 \cdot N_{pixels}) \times(2 \cdot N_{DM})$\bigskip\\
	$\Lambda = \Gamma$ & $(2 \cdot N_{pixels}) \times(2 \cdot N_{DM})$\bigskip\\
	$P_0 = E[(x_0 - \hat x_0)(x_0 - \hat x_0)^T]$ & $(2 \cdot N_{pixels}) \times(2 \cdot N_{pixels})$\bigskip\\
	$Q_k = \Lambda E[w_kw_k^T] \Lambda^T$ & $(2 \cdot N_{pixels}) \times(2 \cdot N_{pixels})$\bigskip\\
	$H_k = diag \left(\begin{bmatrix} \Re \{G_{DM2} \phi_{k1}\} & \Im \{G_{DM2} \phi_{k1}\} \\ \vdots & \vdots \\ \Re \{G_{DM2} \phi_{kj}\} & \Im \{G_{DM2} \phi_{kj}\} \end{bmatrix}_{pixel} \right) $ &  $(N_{pixels})\cdot(N_{pairs}) \times(2 \cdot N_{pixels}) $\bigskip\\
	$R_k = E[n_k n_k^T]$ & $(N_{pixels})\cdot (N_{pairs}) \times (N_{pixels})\cdot(N_{pairs}) $ \bigskip\\
	$K_k$ is computed & $(2 \cdot N_{pixels})  \times(N_{pixels})\cdot(N_{pairs})$\bigskip\\
  \bottomrule
   \end{tabular}\vline
   \caption{Definition of all filter Matrices.  $N_{DM}$ is the number of actuators on a single DM, $N_{pixels}$ is the number of pixels in the area targeted for dark hole generation, and $N_{pairs}$ is the number of image pairs taken while applying positive and negative shapes to the deformable mirror}
   \label{filtertable}
\end{table}

A fundamental property of the Kalman filter is that the optimal gain, Eq.~\ref{eq:gain}, is not based on measurements, but rather estimates of the state covariance, $P_k(-)$, process noise from the actuation $Q_{k-1}$, and sensor noise $R_k$. This means that the optimality of the estimate is closely related to the accuracy and form of these matrices; this will be discussed in \S\ref{sec:noise}. The gain matrix, $K_k$, is ultimately what balances uncertainty in the prior state estimate against uncertainty in the measurements $z_k$ when computing the final state estimate update, $\hat x_k(+)$.  

\begin{table}[ht!]
   \centering
   \vline
   \begin{tabular}{ cc } 
      \toprule
      Variable & Dimension \\
      \midrule
   	$z = \begin{bmatrix} 
		\begin{bmatrix} I_1^+ - I_1^- \\ \vdots \\ I_j^+ - I_j^- \end{bmatrix}_{1} \\ 
		\vdots \\  
		\begin{bmatrix} I_1^+ - I_1^- \\ \vdots \\ I_j^+ - I_j^- \end{bmatrix}_{n} 
		\end{bmatrix}$ & $(N_{pixels})\cdot (N_{pairs}) \times 1$\bigskip\\
   	$\hat x = \begin{bmatrix} 
			\begin{bmatrix} \Re \{E_1\} \\ \Im\{E_1\} \end{bmatrix} \\ 
			\vdots \\ 
			\begin{bmatrix} \Re \{E_n\} \\ \Im\{E_n\} \end{bmatrix}
			\end{bmatrix}$  & $(2 \cdot N_{pixels}) \times 1$ \bigskip\\
	$u = \begin{bmatrix} DM1 \\ DM2 \end{bmatrix}$ & $ (2 \cdot N_{DM}) \times 1$\bigskip\\
	$w = \begin{bmatrix} \sigma_{DM1} \\ \sigma_{DM2} \end{bmatrix}$ & $(2 \cdot N_{DM}) \times 1$\bigskip\\
	  \bottomrule
 \end{tabular}\vline
 \caption{Definition of Filter Vectors.  $N_{DM}$ is the number of actuators on a single DM, $N_{pixels}$ is the number of pixels in the area targeted for dark hole generation, and $N_{pairs}$ is the number of image pairs taken while applying positive and negative shapes to the deformable mirror}
 \label{filtervectors}
 \end{table}

$H_k$ is constructed by separating the real and imaginary parts of the DM probe field. Thus it will be underdetermined unless at least $2$ pairs of images are used in the measurement, one of the major limitations of the DM Diversity algorithm. This will result in a non-unique solution to the state when using a batch-process, and will only provide the solution with the smallest quadratic norm since it must be solved via the right pseudo-inverse. On the other hand, the Kalman filter only requires a single measurement  as an update to the state. Therefore it isn't necessary for the matrix to be square or overdetermined, and we maintain a favorable dimensionality when updating the state. 

\section{Iterative Kalman Filter}\label{sec:iterative}
An additional advantage of the Kalman filter is that we may apply the filter iteratively, feeding the newly computed state $\hat x_K(+)$ and covariance update $P_k(+)$ back into the filter again, setting $u_{k-1}$ to zero. For sufficiently small control this will help account for nonlinearity in the actuation and better filter noise in the system, limited only by the accuracy of the observation matrix, $H_k$. With no control update, the control signal will be set to zero when we iterate the filter. Following a notation similar to Gelb \cite{gelb1974optimal}, the $j^{th}$ iteration of feedback into the iterative Kalman filter at the $k^{th}$ control step is
\begin{align}
\hat x_{j,k} (-) &=  \hat x_{j-1,k-1}(+) \label{eq:iterstate}\\
P_{j,k}(-) &= \Phi_{j-1,k-1} P_{j-1,k-1}(+) \Phi_{j-1,k-1}^T + Q_{j-1,k-1} \label{eq:itercov}\\
K_{j,k} &= P_{j,k}(-)H_{j,k}^T \left[H_{j,k}P_{j,k}(-)H_{j,k}^T + R_{j,k} \right]^{-1}\label{eq:itergain}\\
\hat x_{j,k}(+) &= \hat x_{j,k}(-) + K_{j,k} \left[z_{j,k} - H_{j,k} \hat x_{j,k}(-) \right] \label{eq:iterupdate}\\
P_{j,k}(+) &= \left[P_{,j}k(-)^{-1} + H_{j,k}^T R_{j,k}^{-1}H_{j,k} \right]^{-1}.\label{eq:itercov}
\end{align}

The power of iterating the filter lies in what we are fundamentally trying to achieve. For a successful control signal, we will have suppressed the field. This means that the magnitude of the probe signal will be lower than the control perturbation. This guarantees that $H_k$ will better satisfy the linearity condition than $\Gamma u$. As a result, if we iterate the filter on itself during a given control step we can use the discrepancy between the image predicted by $H_k \hat x_k(+)$ and the measurements, $z_k$, in Eq.~\ref{eq:iterupdate} to filter out any error due to nonlinear terms not accounted for in $\Gamma$. In this way, we can accommodate a small amount of nonlinearity in our extrapolation of the state without having to resort to a nonlinear, or extended, Kalman filter. This means that we don't have to re-linearize about $\hat x_k(+)$, as would be the case for an iterative extended Kalman filter (IEKF). It also avoids having to concern ourselves with any bias introduced into the estimate by a nonlinear filter. It should be pointed out here that while we have chosen not to in this case, we can move to an IEKF by simply re-linearizing about the estimate output, $\hat x_{j-1,k}(+)$, at each step $j$ until the estimate converges.
\section{Sensor and Process Noise}\label{sec:noise}
Two important design parameters for the performance of the filter are the process noise, $Q_{k-1}$, and the sensor noise, $R_k$. In order for the filter to operate optimally in the laboratory we make reasonable assumptions by appealing to physical scaling of the two largest known sources of error in the system. Our sensor noise will be determined by the dark current and read noise inherent to our detector. Our process noise will largely come from errors in our actuation shape. 
 
We treat the process noise as poor knowledge of the DM surface, which comes from the inherent nonlinearity in the voltage-to-actuation gain as a function of voltage, the variance in this gain from actuator to actuator, and the accuracy of the superposition model used to construct the mirror surface that covers the 32x32 actuator array of the Boston Micromachines kilo-DM. Physical models, such as those found in Blain et al. \cite{blain2010dm}, have been constructed to produce a more accurate surface prediction over the full $1.5 \mu m$ stroke range with an rms error of $\approx 10$ nanometers. The Kalman filter presents a solution where we can treat actuation errors as additive process noise and include them in the estimator in a statistical fashion, rather than deterministically in a physical model. Since there is no physical reasoning to justify varying $Q_k$ at each iteration, it will be kept constant throughout the entire control history ($Q_{k-1} = Q =$ constant). Two versions of process noise can be considered in this case. The first is where there is no correlation between actuators, giving a purely diagonal matrix with a magnitude corresponding to the square of the actuation variance, $\sigma_u$. The second version of $Q$ which we may consider is one that has symmetric off-diagonal elements. This will treat uncertainty due to inter-actuator coupling and errors in the superposition model statistically. As a first step, we will not consider inter-actuator coupling to help avoid a poorly conditioned matrix. This helps guarantee that the Kalman filter itself will be well behaved. Thus the process noise for the filter will be

\begin{equation}
Q = \sigma_{u}^2 \Gamma \mathcal I  \Gamma^T.
\end{equation}

Following Howell \cite{howell2000handbook} the noise from both the incident light and dark noise in a CCD detector follows a Poisson distribution. Since our measurement is a difference of pairwise images each measurement will follow a zero-mean distribution that will become more Gaussian as the exposure time increases. We simplify the noise statistics by assuming it is uncorrelated and constant from pixel to pixel.  $R_k$ is thus a diagonal matrix of the mean pixel covariance, $\sigma_{CCD}$, given by
\begin{equation}
R = \frac{\sigma_{CCD}}{I_{00}} \mathcal I_{N_{pairs} \times N_{pairs}},\label{eq:sensnoise}
\end{equation}

where $I_{00}$ is the peak count rate of the PSF's core (allowing us to describe $R$ in units of contrast). Having appealed to physical scaling in the HCIL, we now have close approximations of the true process and sensor noise exhibited in the experiment.
\section{Optimal Probes: Using the Control Signal}\label{sec:optprobes}
The choice of the probe shape used for estimation is critical to create a well-posed problem. These shapes are typically chosen to be analytical functions that can be proven analytically to modulate the field as uniformly as possible. However, there is no true formalism to determine the ``best" shapes to probe the dark hole. In any dark hole there are discrete aberrations that are much brighter than others, requiring that we apply more amplitude to those spatial frequencies. Conversely the bright speckles raise the amplitude of the probe shape, making it too bright for to take a good measurement of dimmer speckles. Excluding this issue, we also cannot truly generate the analytical functions, meaning that we could get a poor measurement update. Even the DM with the highest actuator density available, the Boston Micromachines 4K-DM\copyright, can only approximate each function with 64 actuators. We account for the true shape in our model but this shape does not truly probe each pixel in the dark hole with equal weight, which was the primary advantage of the analytical function for a probe shape in the first place. Fortunately, we can once again appeal to our mathematical model for estimation and control to help determine an adequate probe shape. In closed loop, the control law determines a shape to suppress the speckles in the dark hole. First, we recognize that the controller has necessarily modulated the aberrated field in the dark hole. In principle, if we apply the conjugate of the control shape we will increase the energy of the aberrated field. Instead of applying probes in addition to the control shape, we use the control shape itself (and its negative) to probe the elecric field in the dark hole. Thus, we can rely on the controller to compute probe shapes, eliminating our reliance on an analytical form. In this way, we can optimally choose probe shapes that will inherently modulate brighter speckles more strongly than dimmer speckles and we have confidence that the shape will adequately perturb the dark hole field.
\section{Experimental Results for the Kalman Filter Estimator in Monochromatic Light}
\begin{figure}[h!]
\centering
\subfigure[]{\includegraphics[width = 0.32\textwidth]{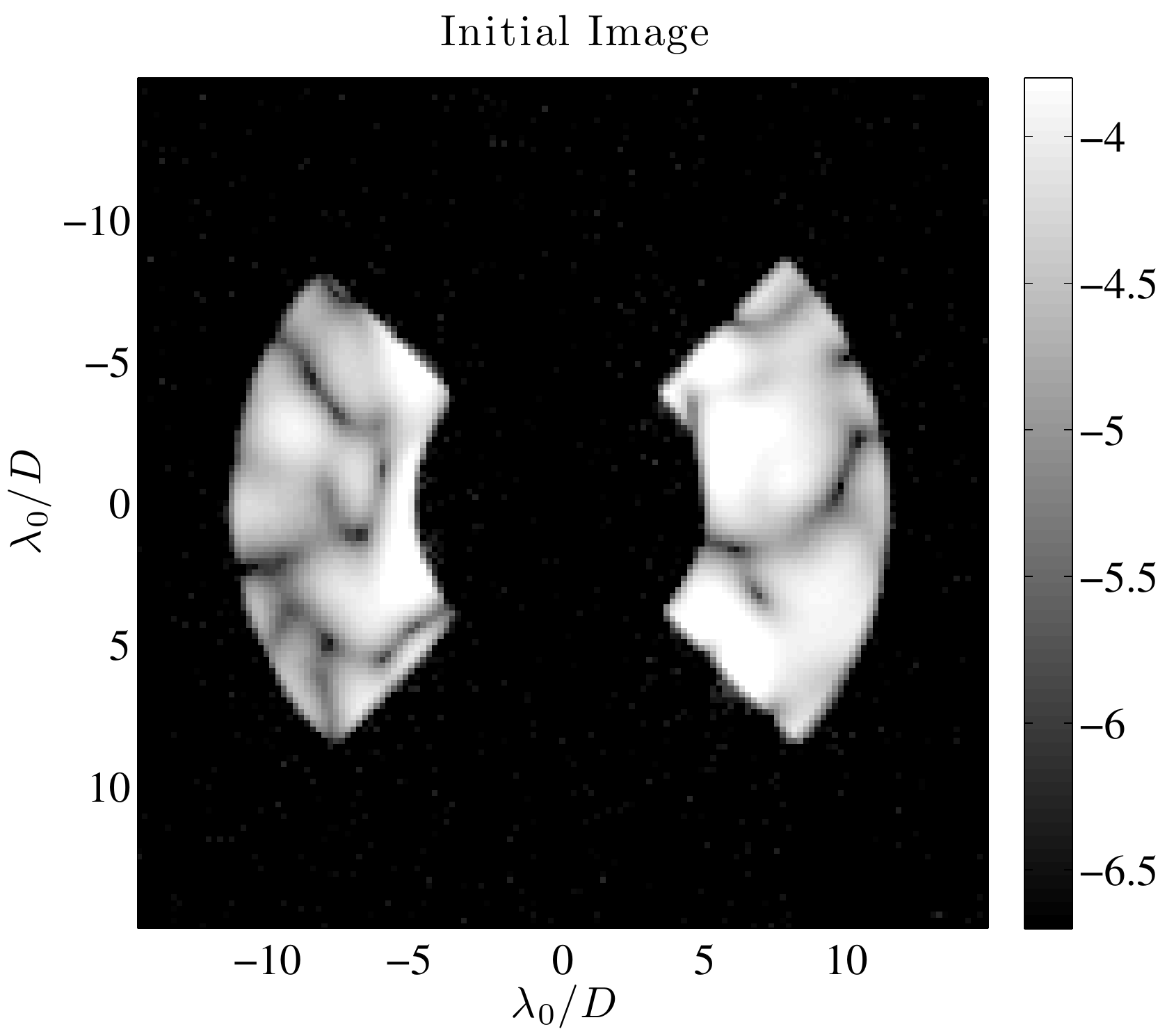}\label{fig:4pairs_initial}}
\subfigure[]{\includegraphics[width = 0.335\textwidth]{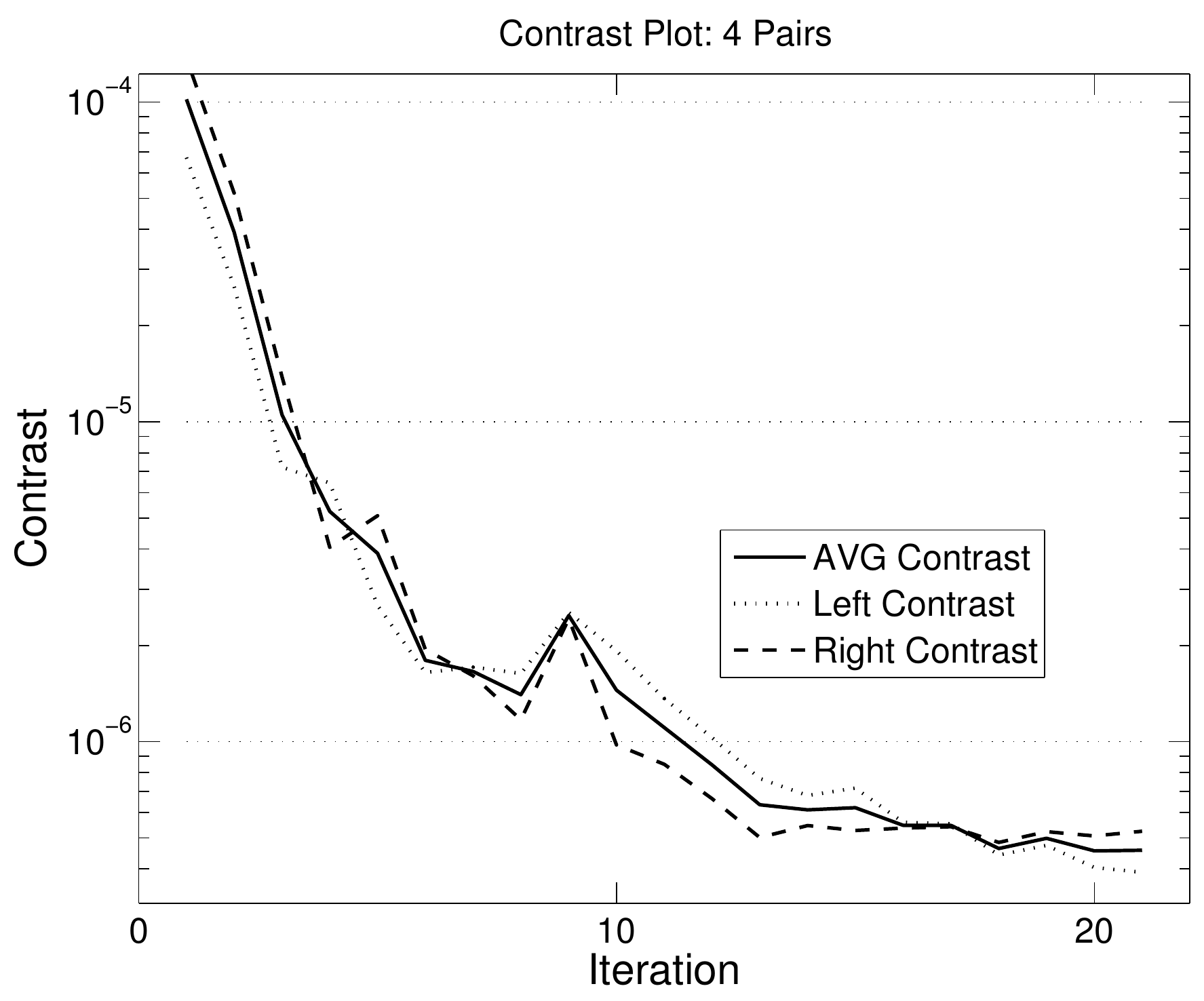}\label{fig:4pairs_contrast}}
\subfigure[]{\includegraphics[width = 0.32\textwidth]{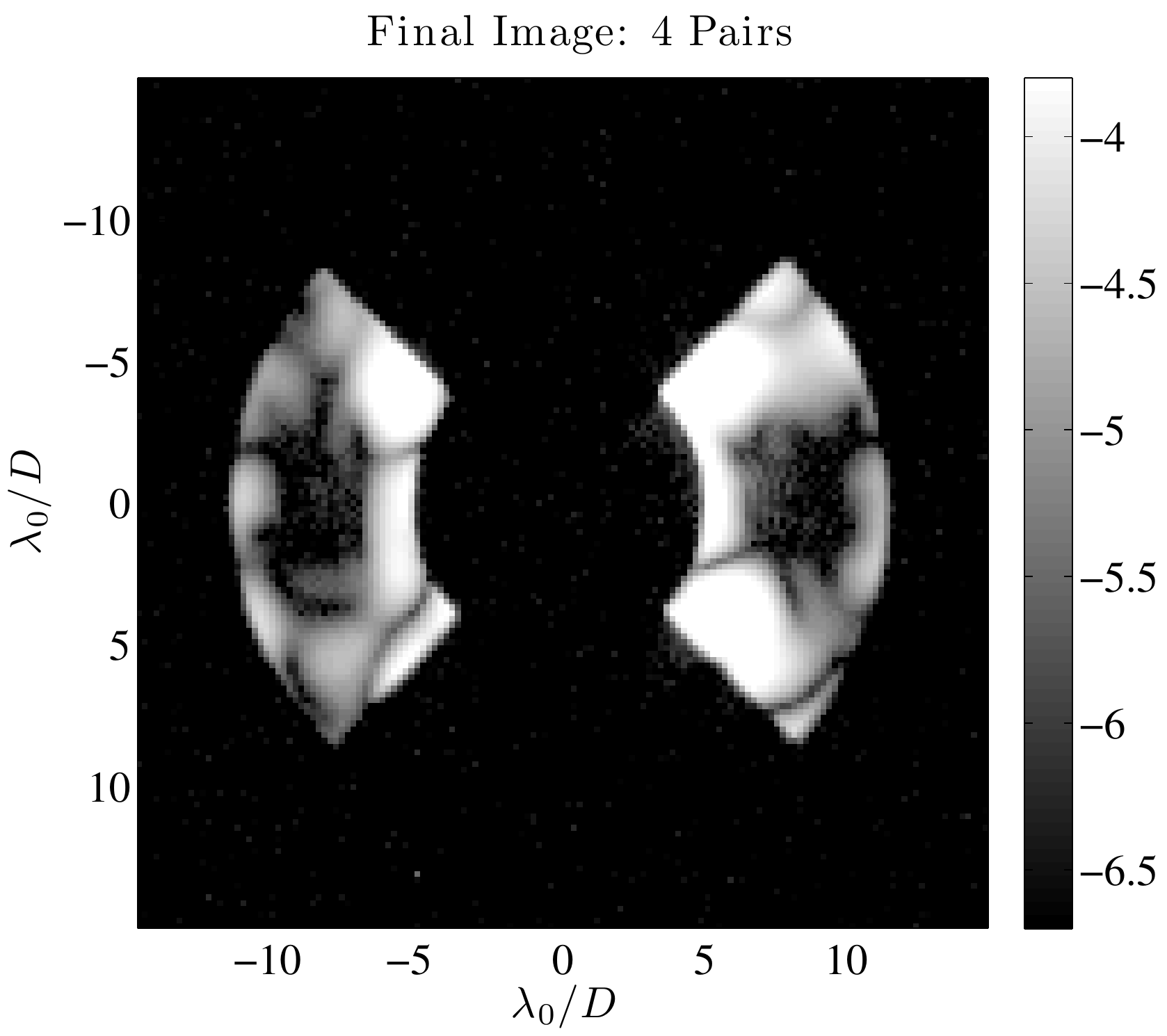}\label{fig:4pairs_final}}
\caption{Experimental results of sequential DM correction using the discrete time extended Kalman filter with 4 image pairs to build the image plane measurement, $z_k$.  The dark hole is a square opening from 7--10 $\times$ -2--2 $\lambda/D$ on both sides of the image plane.  (a) The aberrated image.  (b)  Contrast plot. (c) The corrected image. Image units are log(contrast). }\label{fig:4pairs}
\end{figure}

We corrected the field using the Kalman filter estimator using four, three, two, and one pair of images as a measurement update to assess the degradation in performance as information is lost. We begin with four measurements (four image pairs), to compare its performance using the same number of measurements as were used in the DM Diversity estimator. Using 4 pairs, the filter achieved a contrast of $4.0 \times 10^{-7}$ in (7-10)x(-2-2) $\lambda/D$ symmetric dark holes within 20 iterations of the controller, shown in Fig.~\ref{fig:4pairs}. Note that this used a total of $160$ estimation images, which is the same amount of information available to the DM Diversity estimator when it achieved a contrast of $3.5 \times 10^{-7}$ in $20$ iterations.

\begin{figure}[h!]
\centering
\subfigure[]{\includegraphics[width = 0.32\textwidth]{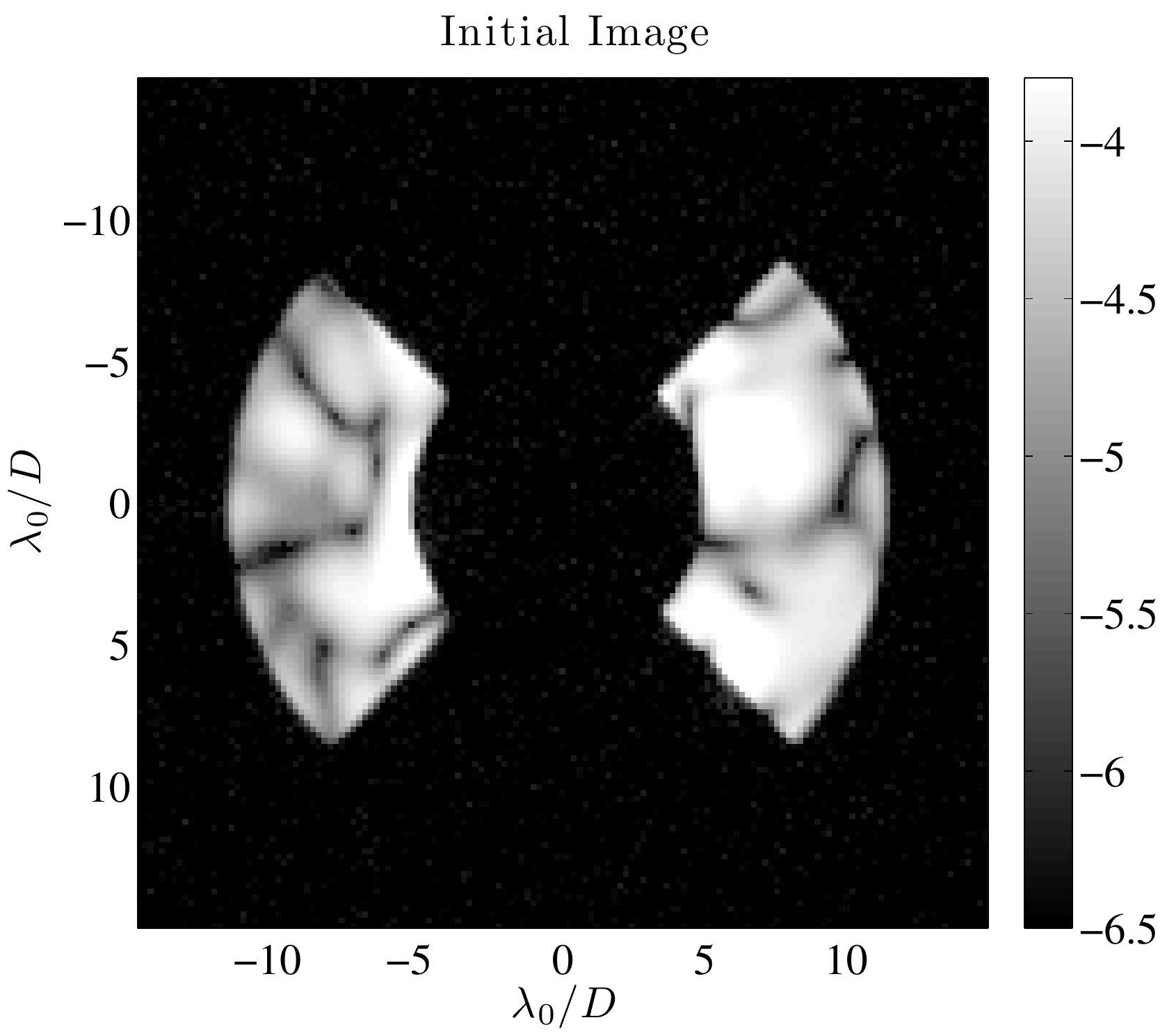}\label{fig:3pairs_initial}}
\subfigure[]{\includegraphics[width = 0.335\textwidth]{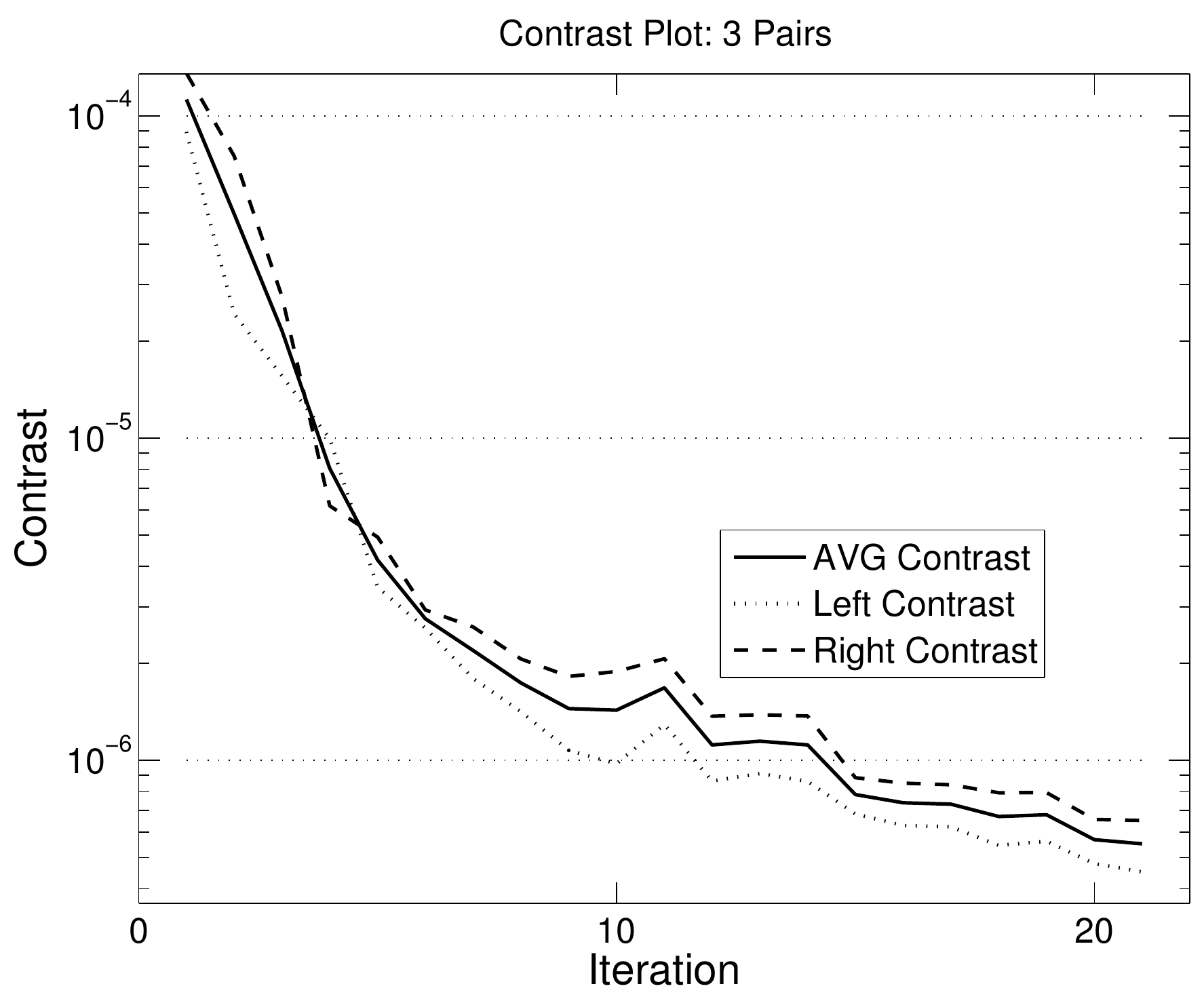}\label{fig:3pairs_contrast}}
\subfigure[]{\includegraphics[width = 0.32\textwidth]{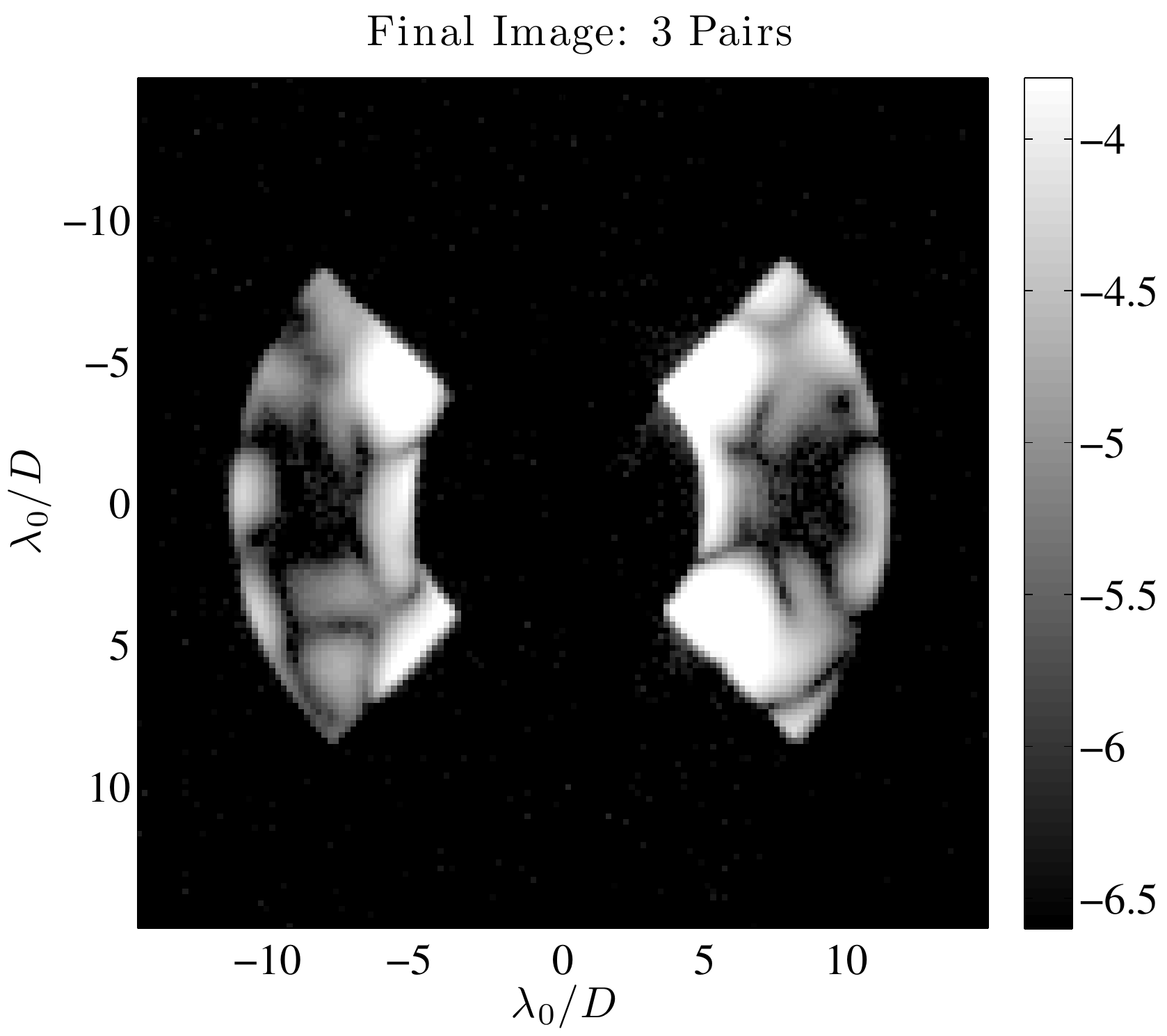}\label{fig:3pairs_final}}
\caption{Experimental results of sequential DM correction using the discrete time extended Kalman filter with 3 image pairs to build the image plane measurement, $z_k$.  The dark hole is a square opening from 7--10 $\times$ -2--2 $\lambda/D$ on both sides of the image plane.  (a) The aberrated image.  (b)  Contrast plot. (c) The corrected image. Image units are log(contrast). }\label{fig:3pairs}
\end{figure}

\begin{figure}[h!]
\centering
\subfigure[]{\includegraphics[width = 0.32\textwidth]{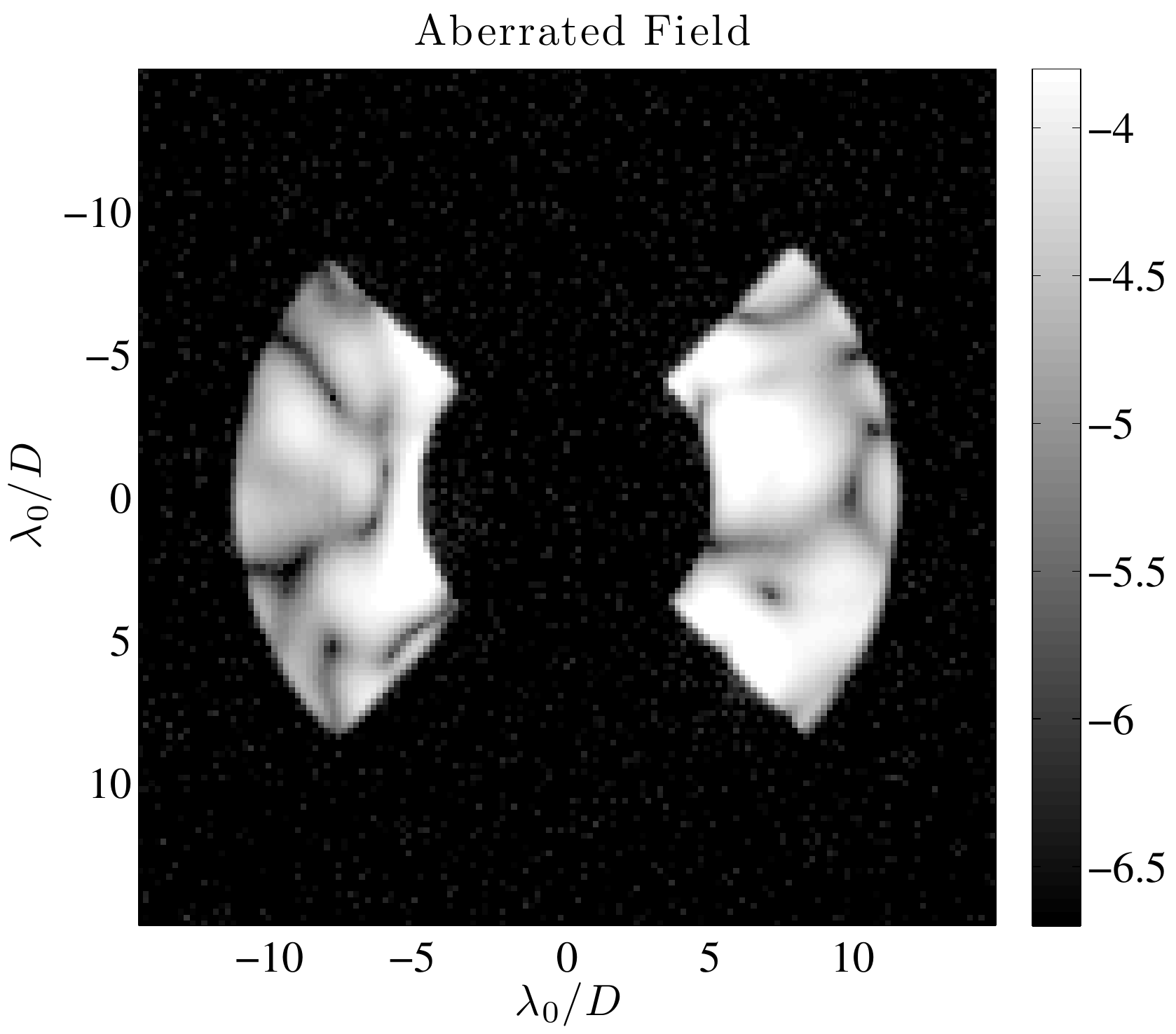}\label{fig:2pairs_initial}}
\subfigure[]{\includegraphics[width = 0.335\textwidth]{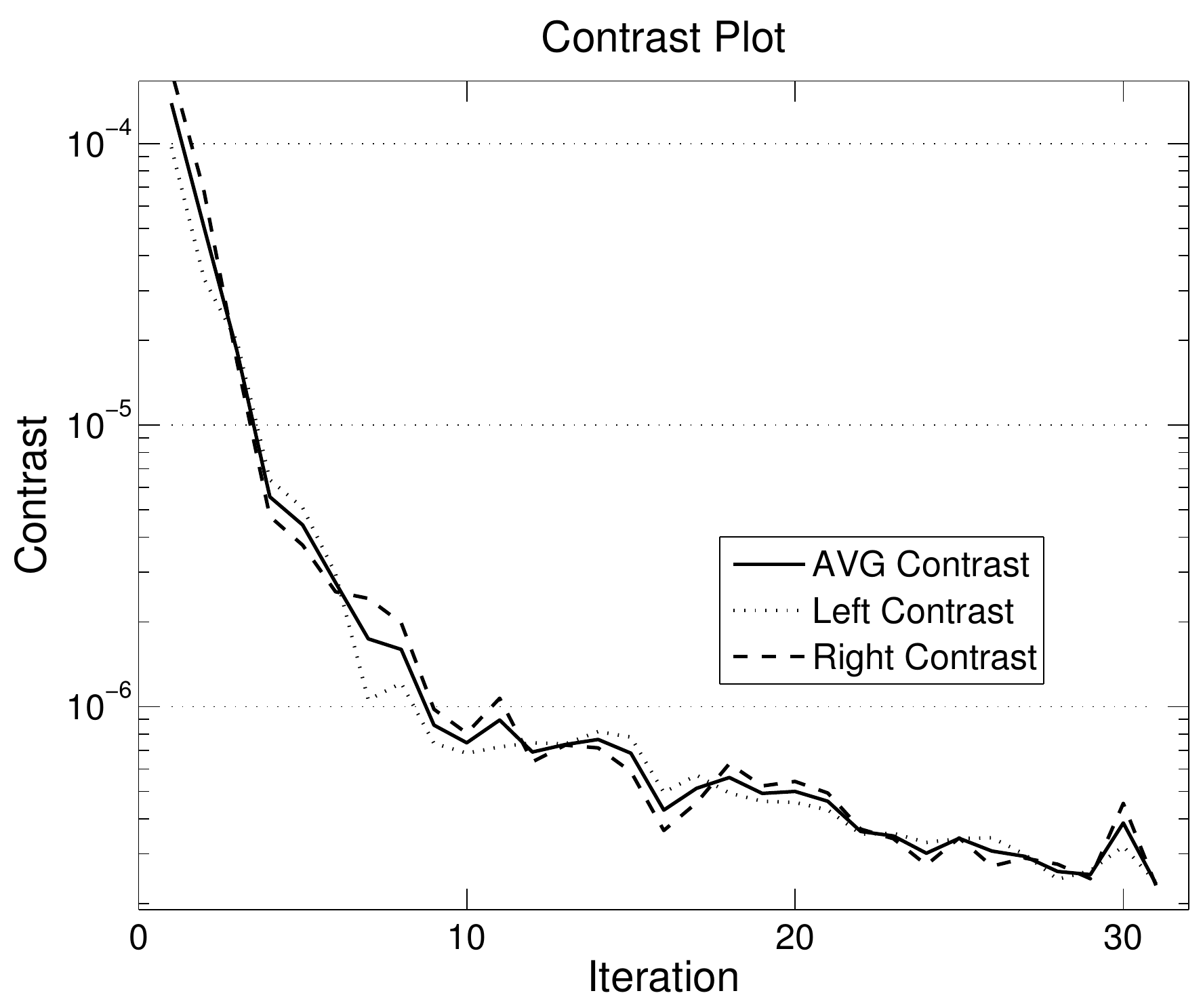}\label{fig:2pairs_contrast}}
\subfigure[]{\includegraphics[width = 0.32\textwidth]{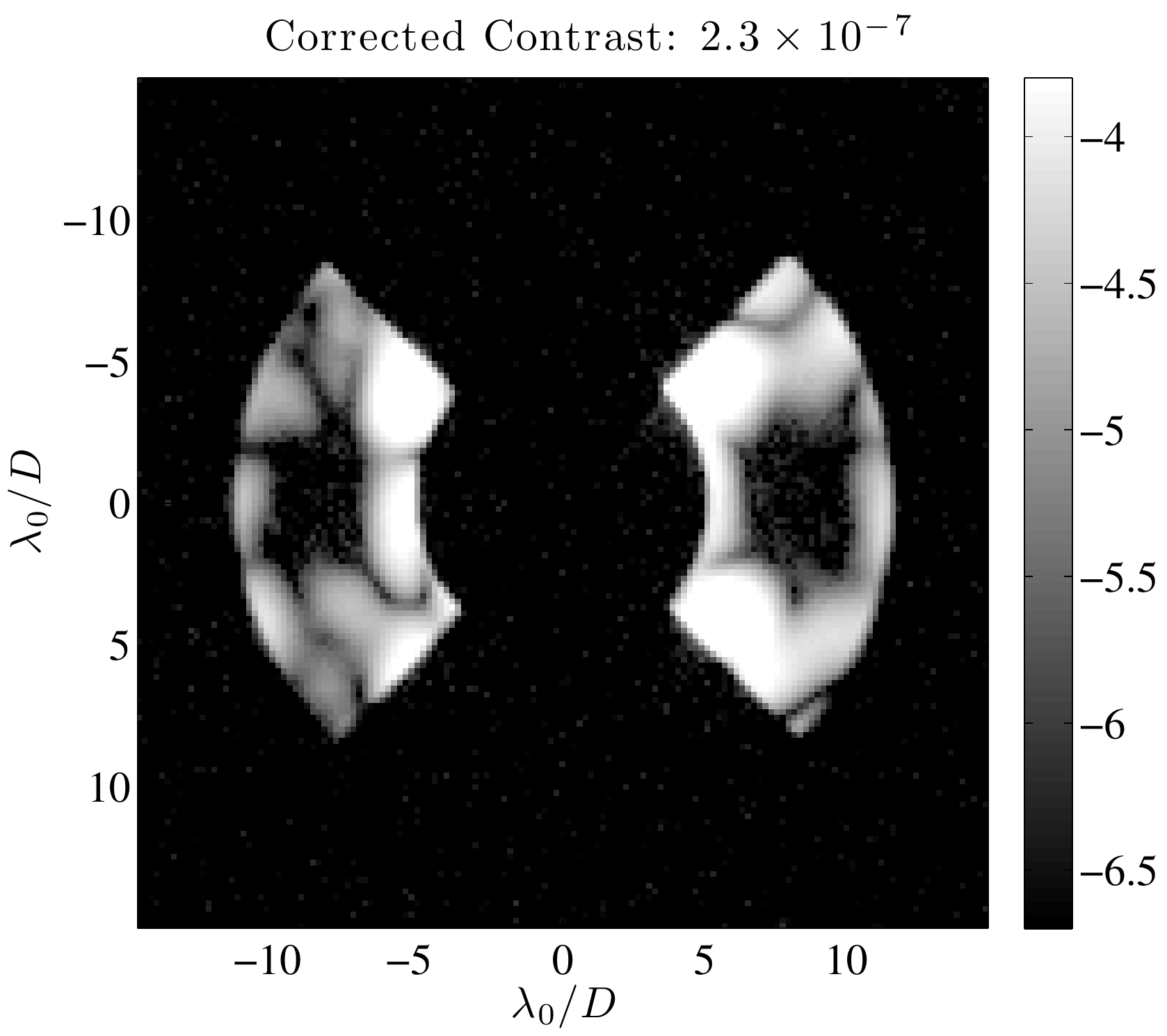}\label{fig:2pairs_final}}
\caption{Experimental results of sequential DM correction using the discrete time extended Kalman filter with 2 image pairs to build the image plane measurement, $z_k$.  The dark hole is a square opening from 7--10 $\times$ -2--2 $\lambda/D$ on both sides of the image plane.  (a) The aberrated image.  (b)  Contrast plot. (c) The corrected image. Image units are log(contrast). }\label{fig:2pairs}
\end{figure}

When the number of image pairs is reduced to three, the correction algorithm was still able to reach a contrast level of $5.0 \times 10^{-7}$ using only $120$ estimation images, as shown in Fig.\ref{fig:3pairs}.
Having proven that we can successfully reach very close to the same limits with fewer exposures, we now tune the covariance initialization and noise matrices and attempt only using two pairs of images. By reducing the number of image pairs to two, we are using half as many images as the correction using the DM Diversity estimator and have reached a point where the batch process method will no longer provide a solution that takes advantage of the averaging effect of the left pseudo-inverse solution. After further tuning the covariance and noise matrices, the contrast achieved after $30$ iterations of the correction algorithm was $2.3\times10^{-7}$, shown in Fig.~\ref{fig:2pairs}. Note that this is better than the case which used three pairs because we have improved the covariance initialization and increased the number of times the filter is iterated in a single control step. In fact it should be noted that making the filter iterative is critical to its performance since it accounts for nonlinearity, particularly in the propagation of the control.

\begin{figure}[h!]
\centering
\subfigure[]{\includegraphics[width = 0.32\textwidth]{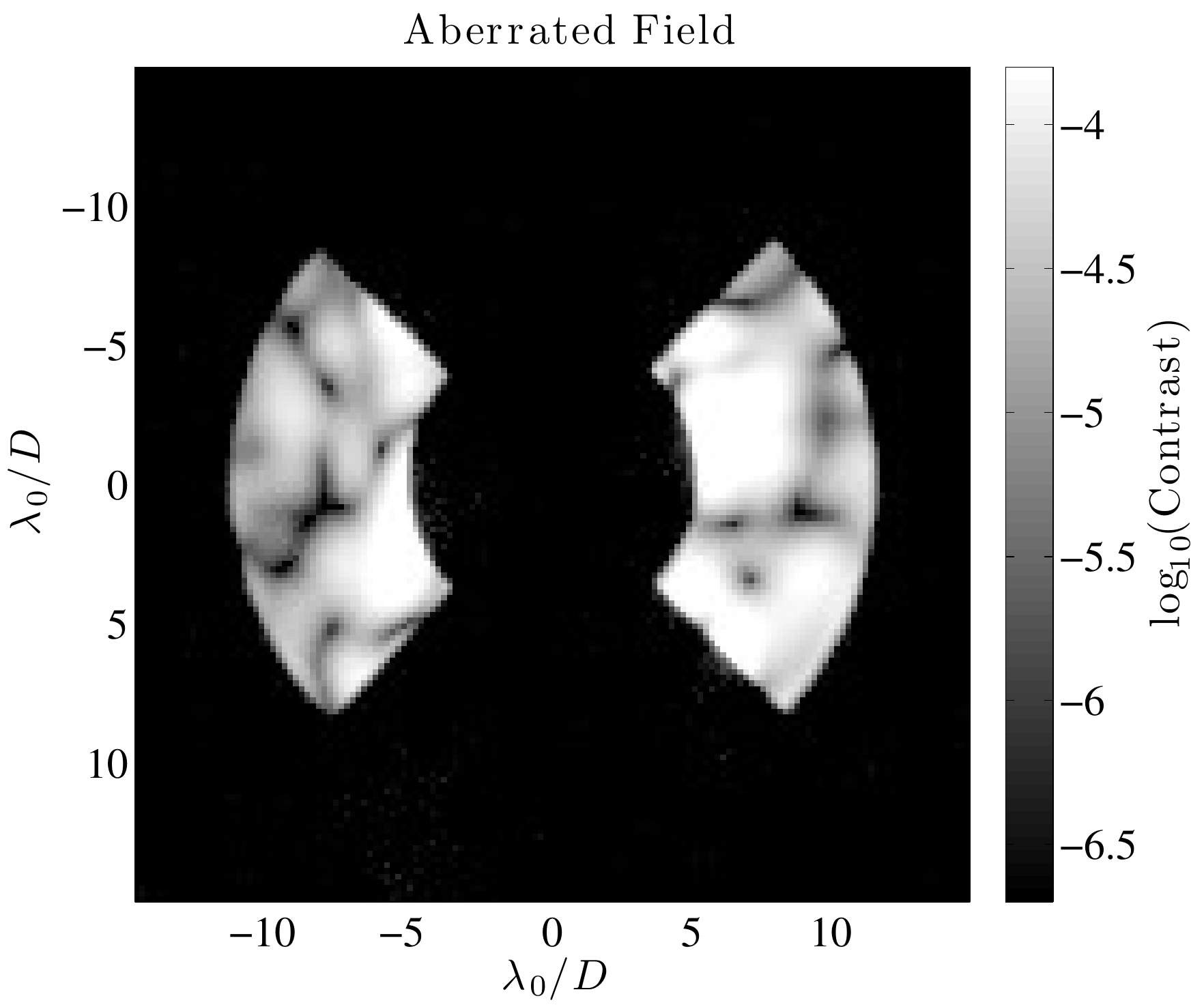}\label{fig:onepair_initial}}
\subfigure[]{\includegraphics[width = 0.335\textwidth]{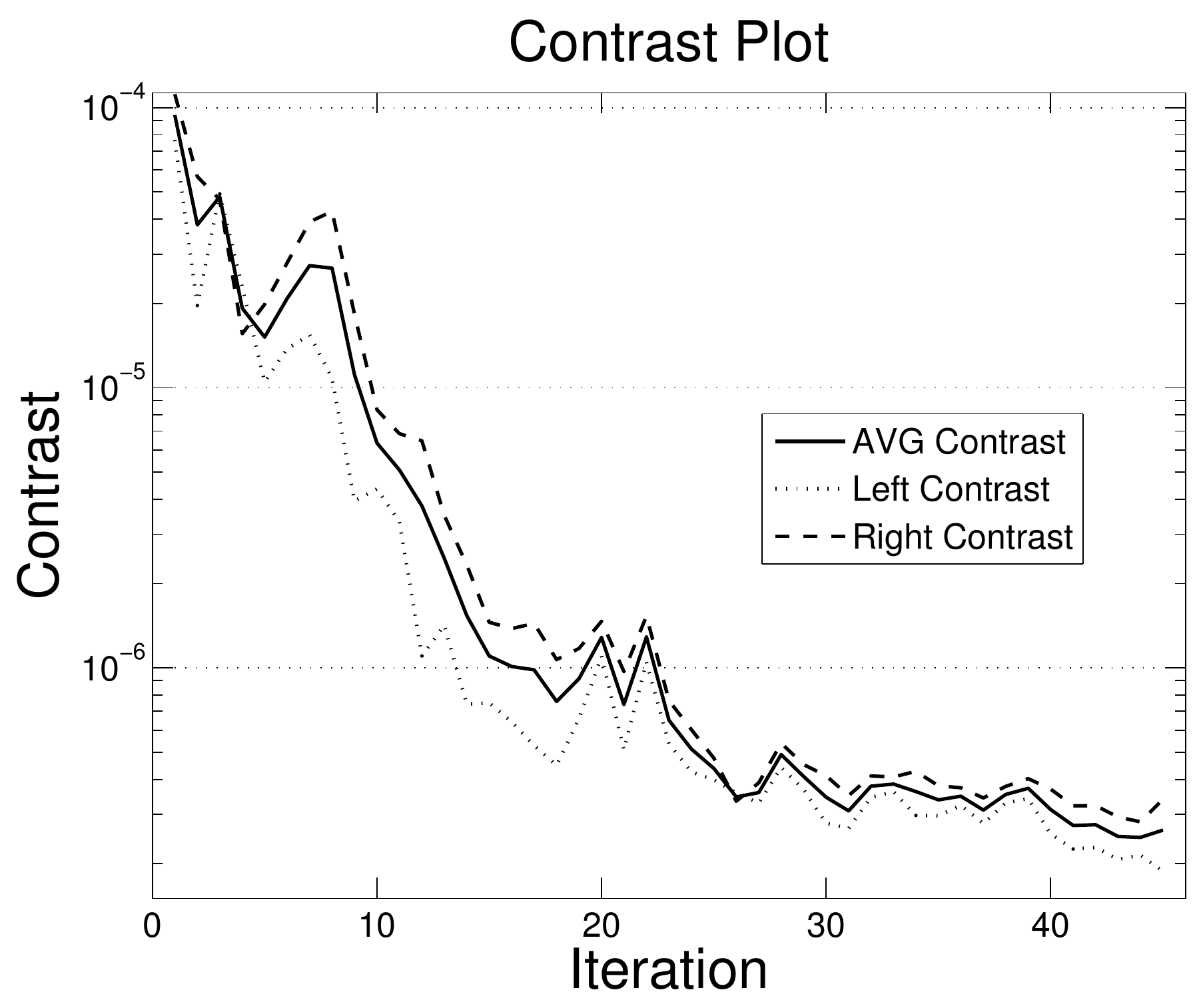}\label{fig:onepair_contrast}}
\subfigure[]{\includegraphics[width = 0.32\textwidth]{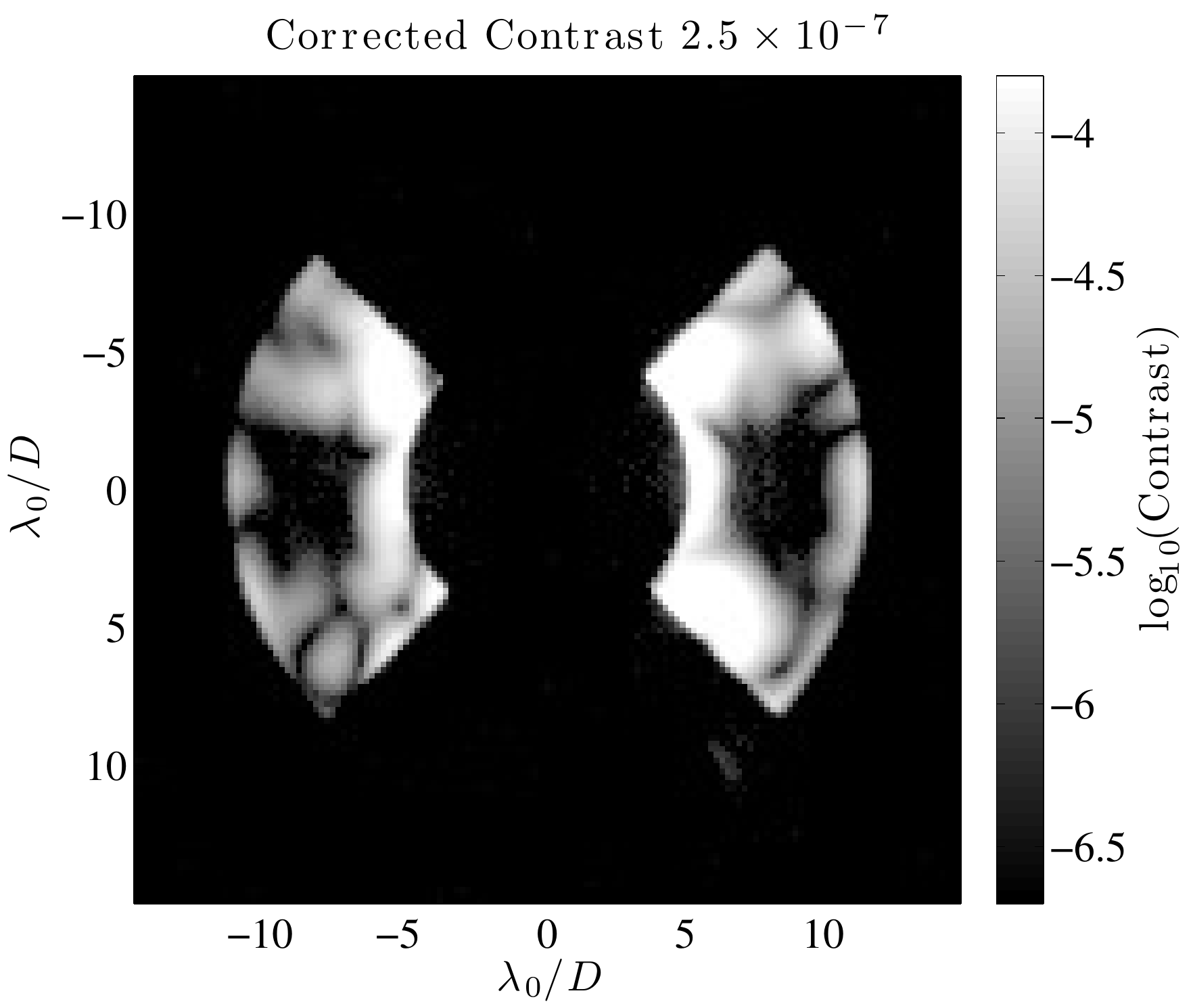}\label{fig:onepair_final}}
\caption{Experimental results of sequential DM correction using the discrete time extended Kalman filter with one image pair to build the image plane measurement, $z_k$.  The dark hole is a square opening from 7--10 $\times$ -2--2 $\lambda/D$ on both sides of the image plane.  (a) The aberrated image.  (b)  Contrast plot. (c) The corrected image. Image units are log(contrast). }\label{fig:1pair}
\end{figure}

Reducing the number of measurements to a single pair we find a very interesting result. The quality of the measurement at any particular time step of the algorithm is now dependent on the quality of that particular probe shape. As a result, if the probe does not happen to modulate the field well the field estimate gets worse. It is also important to cycle through the probe shapes. A single probe may not modulate a specific location of the field well, so we must choose a different probe shape to guarantee that we adequately cover the entire dark hole. Starting from an aberrated field with an average contrast of $9.418\e{-5}$, Fig.~\ref{fig:onepair_initial}, we achieved a contrast of $3.1\e{-7}$ in $30$ iterations and $2.5\e{-7}$ in $43$ iterations of control, Fig.~\ref{fig:onepair_final}. Looking at the contrast plot in Fig.~\ref{fig:onepair_contrast}, the sensitivity of a single measurement update to the quality of the probe is very clear. What is interesting, however, is that the modulation damps out over the control history. While we do not suppress as quickly in earlier iterations, as in the case with more probes, we achieve our ultimate contrast levels in almost the exact same number of iterations. This is a direct result of developing good coverage across the dark hole over time by changing the probe shape at each iteration. Thus, even with one measurement update at each iteration the prior state estimate history stabilizes the estimate in the presence of the measurement update's poor signal-to-noise at high contrast levels. What is further encouraging is that these results use arbitrary probe shapes based on an analytical function, with no criteria to evaluate how effectively each modulates the field. If we were to  choose our probes more carefully, we will see a dramatic improvement in the rate of convergence for a single measurement update.

A very promising aspect of this estimation scheme is that its performance did not degrade significantly as the amount of measurement data was reduced. With only 86 estimation images it was capable of reaching the same final contrast achieved by the DM diversity algorithm in \S\ref{batch}, which achieved a contrast of $2.5\times10^{-7}$ in 30 iterations. The batch process required $240$ images to maintain an estimate of the entire control history, achieving a contrast of $2.3\times10^{-7}$. Thus by making the estimation method more dependent on a model we were able to reduce our need to measure deterministic perturbations in the image plane electric field.

\begin{figure}[h!]
\centering
\subfigure[]{\includegraphics[width = 0.32\textwidth]{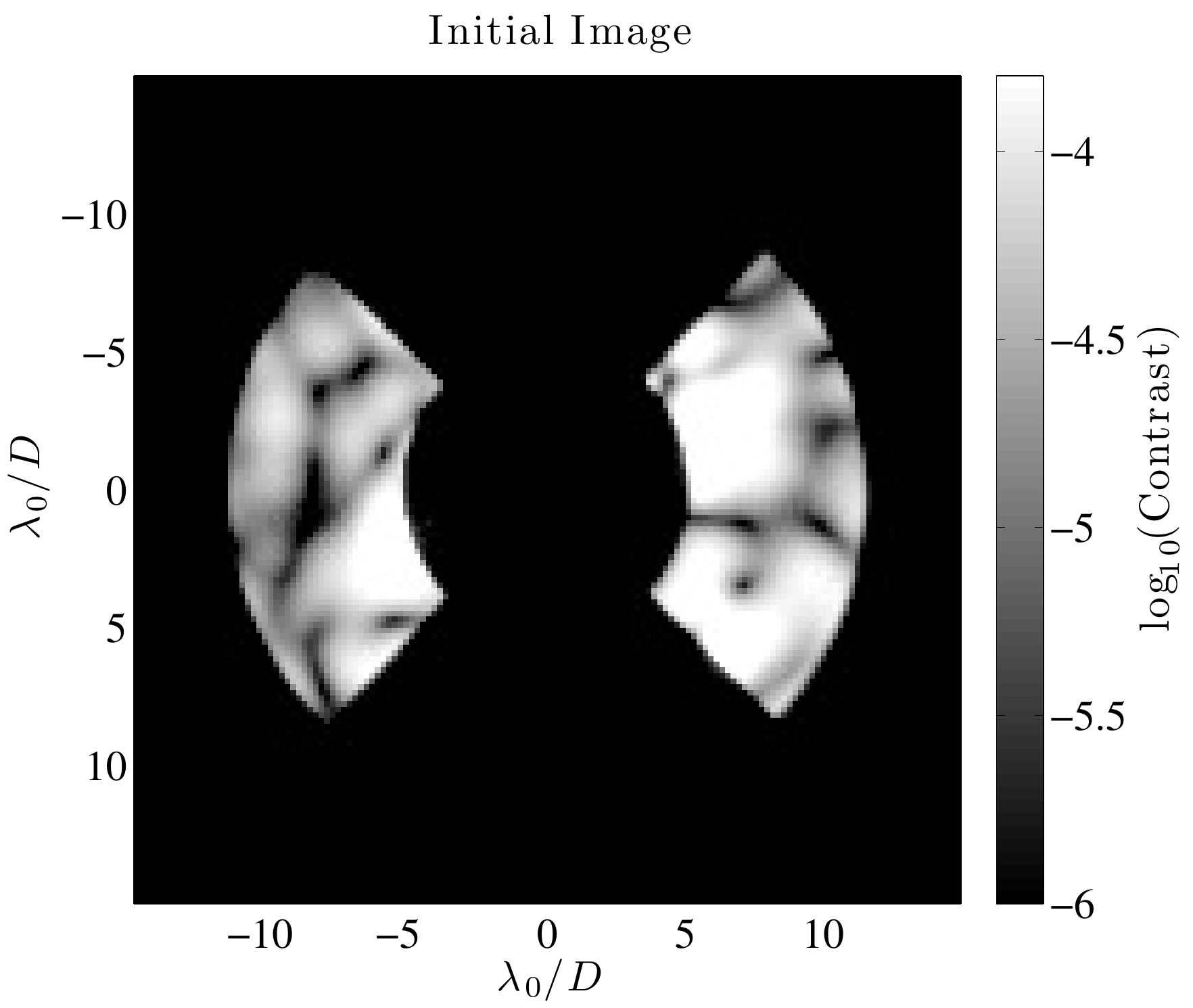}\label{fig:dmprobe_initial}}
\subfigure[]{\includegraphics[width = 0.335\textwidth]{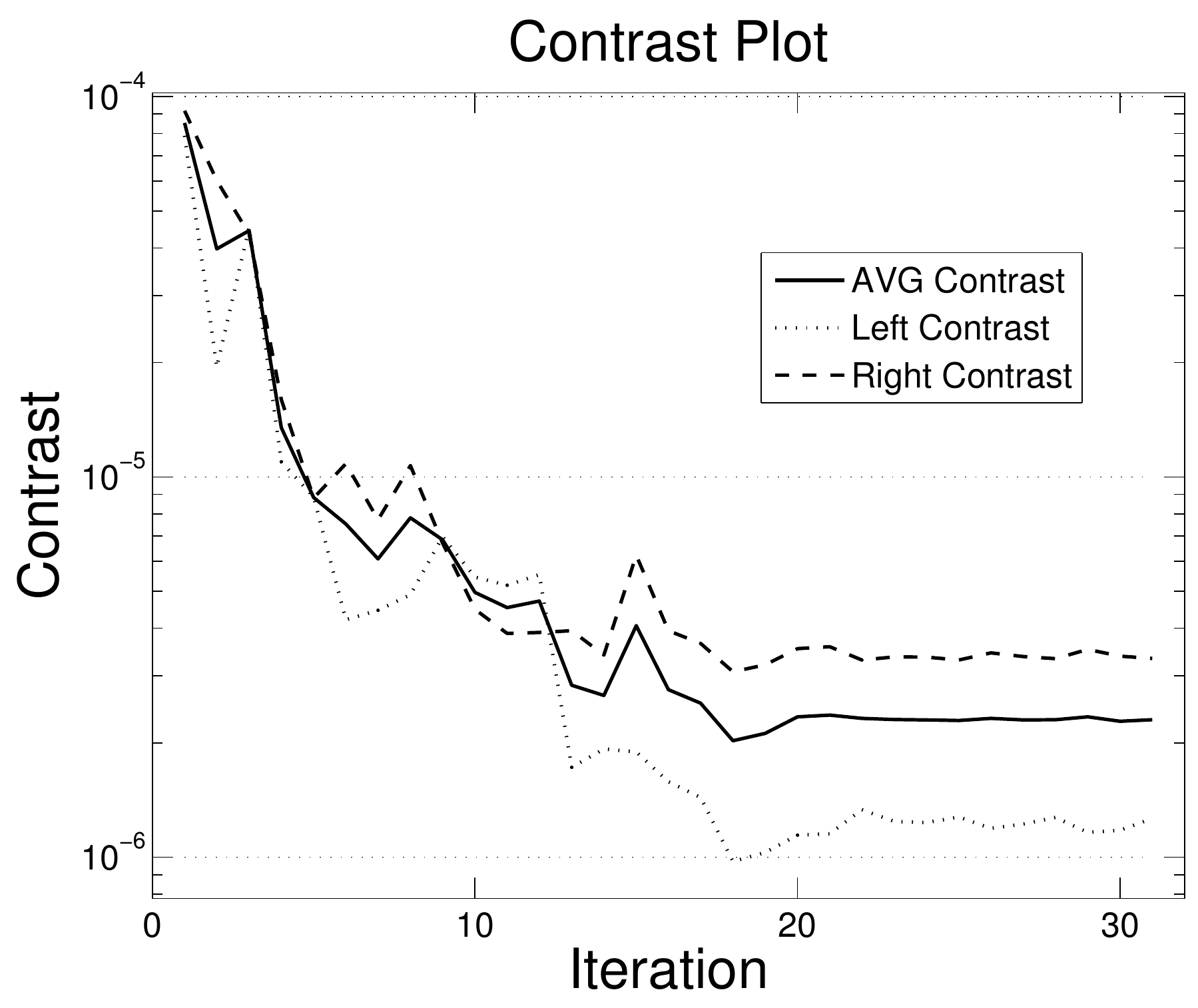}\label{fig:dmprobe_contrast}}
\subfigure[]{\includegraphics[width = 0.32\textwidth]{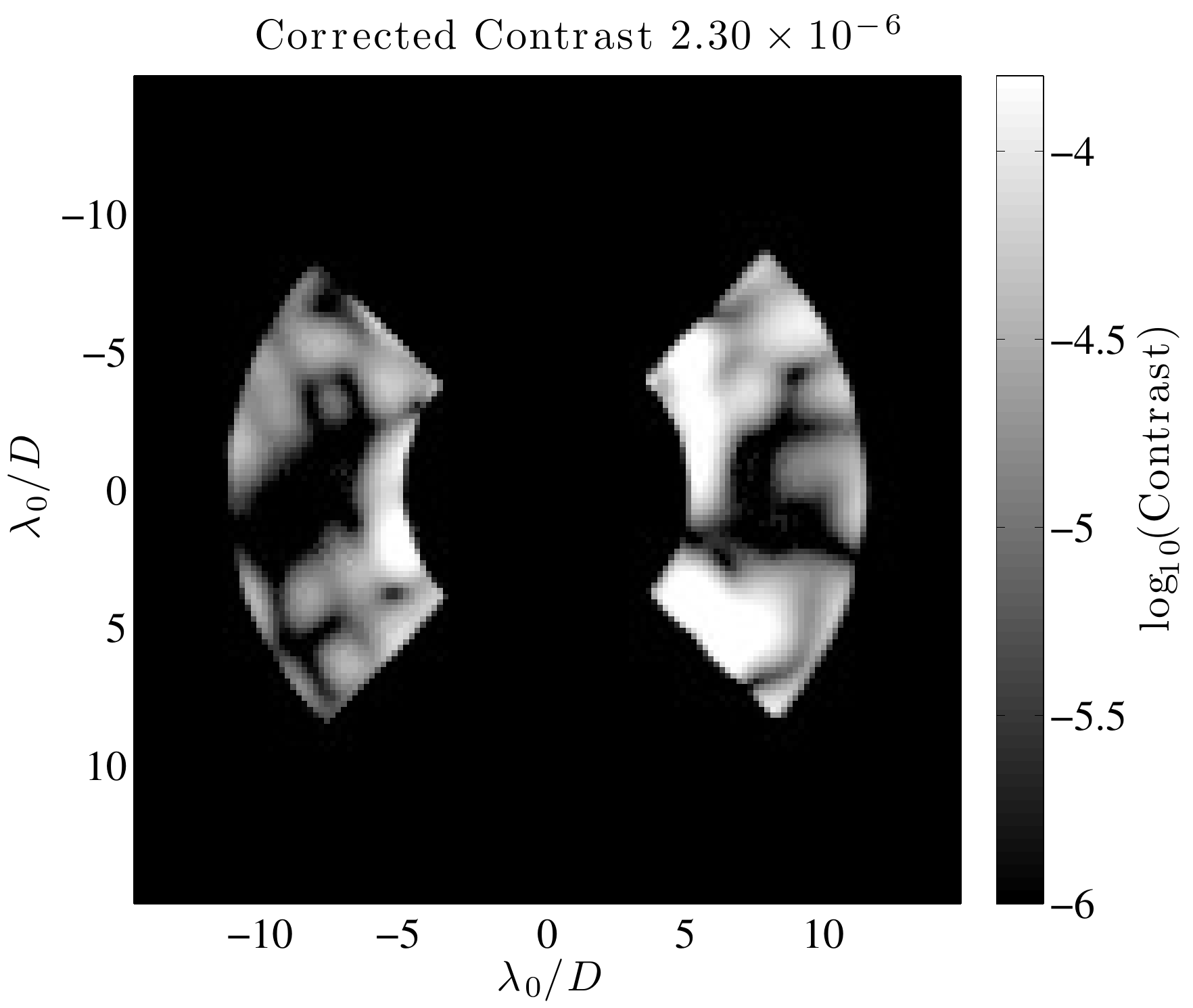}\label{fig:dmprobe_final}}
\caption{Experimental results of sequential DM correction using the discrete time extended Kalman filter with the control effect and it's conjugate shape used as the only probe pair for the measurement update, $z_k$.  The dark hole is a square opening from 7--10 $\times$ -2--2 $\lambda/D$ on both sides of the image plane.  (a) The aberrated image.  (b)  Contrast plot. (c) The corrected image. Image units are log(contrast). }\label{fig:probepair}
\end{figure}

Fig.~\ref{fig:probepair} shows the preliminary results using the control and its conjugate shape as the probe pair for estimating the field with the Kalman filter. The current contrast level is limited to $2.30\e{-6}$ on both sides of the image plane. The assymetry of the dark holes actually hints at the reason for not achieving a higher contrast level. In a 2-DM system, the idea that the optimal control signal can be used as the probe shape requires that we use both mirrors to probe the field. Since the estimator is currently written assuming one mirror is probing the field, we were forced to collapse both control shapes onto the same DM. This means that we are not fully perturbing the field as we would have expected, leaving a particular set of aberrations unprobed. This is not a problem for a single sided dark hole using one DM, where the control and probe surfaces are one and the same. For a two DM system, we simply need to reformulate the estimator as a function of both DMs for this concept to be effective.

\section{Conclusions}\label{conclusions}
In this paper we have demonstrated a discrete time extended Kalman filter to estimate the image plane electric field in closed loop. This type of progress is critical for improving the efficiency of future coronagraphic missions, thereby maximizing the likelihood of planetary detections. We demonstrate the fastest suppression to date in the Princeton HCIL by only requiring a single measurement at each iteration, currently requiring $\approx$30\% of the original set of images for estimation. Faster algorithms also makes focal plane estimation techniques more feasible for ground-based coronagraphic instruments. The closed loop nature of the estimator also provides a more stable to measurement because a measurement update with poor signal-to-noise does not adversely affect the covariance of the state estimate. We have also shown with the single measurement update that not all probe shapes are best for estimation, motivating us to try using the control shape as the probing function. Preliminary results for using the control shape as a probe signal is promising, and may further reduce the number of required exposures to one image per iteration. The Kalman filter also opens up the possibility of adaptive control techniques to learn laboratory physical parameters and bias estimation to gain certainty in planetary detection using only the control history.

\section*{Acknowledgements}
This work was funded by NASA Grant \# NNX09AB96G and the NASA Earth and Space Science Fellowship
\bibliographystyle{spiebib}

\end{document}